\documentclass{aa}
\usepackage{graphicx}
\usepackage{txfonts}
\usepackage{epsfig}

\begin{document}

\def\etal{et al.\ }
\def\kms{km\thinspace s$^{-1}$ }
\def\Lsun{L$_\odot$}

\def\water{H$_2$O~}
\def\Msun{M$_\odot$}
\def\ms{m\thinspace s$^{-1}$}
\def\percc{cm$^{-3}$}

\title{Full polarization study of SiO masers at 86 GHz}

\author{F.\,Herpin\inst{1}, A.\,Baudry\inst{1}, C.\,Thum\inst{2}, D.\,Morris\inst{2}$^,$\inst{3} and H.\,Wiesemeyer\inst{2}} 

\institute{
 Observatoire Aquitain des Sciences de l'Univers, Laboratoire d'Astrodynamique, d'Astrophysique et d'A\'eronomie de Bordeaux, CNRS/INSU UMR n$^{\circ}$ 5804, BP 89, 33270, France
\and
IRAM, 300 rue de la Piscine, Domaine Universitaire, 38406 Saint Martin d'H\`eres, France
\and
Present address: Raman Research Institute, 560080 Bangalore, India
}

\titlerunning{SiO maser polarization}

\abstract{We study the polarization of the SiO maser emission in a representative sample of evolved stars in order to derive an estimate of the strength of the magnetic field, and thus determine the influence of this magnetic field on evolved stars. We made simultaneous spectroscopic measurements of the 4 Stokes parameters, from which we derived the circular and linear polarization levels. The observations were made with the IF polarimeter installed at the IRAM 30m telescope. A discussion of the existing SiO maser models is developed in the light of our observations. Under the Zeeman splitting hypothesis, we derive an estimate of the strength of the magnetic field. The averaged magnetic field varies between 0 and 20 Gauss, with a mean value of 3.5 Gauss, and follows a $1/r$ law throughout the circumstellar envelope. As a consequence, the magnetic field may play the role of a shaping, or perhaps collimating agent of the circumstellar envelopes in evolved objects.}

\maketitle 


\section{Introduction}

The prodigious mass loss observed in numerous and widespread evolved stars make these objects the main recycling agents of the interstellar medium, and thus one of the most important objects in the Universe. Even though our knowledge of evolved stars has considerably improved over recent years, some of their main characteristics remain insufficiently understood (see the review by Herwig 2003):  which mechanisms are responsible for their drastic change of geometry when evolving to the Planetary Nebula (hereafter PN) stage ? What is powering so efficiently the mass loss and could the magnetic field play a major role ? 

 Important information about the physics and chemistry prevailing in the circumstellar envelope  (hereafter {\em CSE}) of evolved stars can be retrieved from radiowave line emission of molecules, specially from maser emission (see the review by Bujarrabal 2003). These envelopes can be probed at different depths through the study of three masing molecules, OH, \water and SiO. Our current knowledge indicates that:
\begin{itemize}
  \item OH radiation traces the outer part of the envelope, at 1000-10000 AU from the central star;
  \item \water molecules are located at intermediate distances, i.e. a few 100 AU;
  \item SiO maser emission comes from the inner regions of the envelope, between 5 to 10 AU (a few stellar radii R$_{\star}$).
\end{itemize}

The SiO maser emission is produced in small gas cells, and  is known to be polarized. The polarization (circular or linear) and angle of the emission can be measured and thus improve our knowledge of these objects. In addition, studying the maser polarization can shed light on the maser theory itself. As explained further in this paper, several uncertainties in the theory make data interpretation often difficult, and new observational data are helpful. One of the most interesting quantities that can be derived from polarization measurements is the stellar magnetic field. According to theory (e.g. Elitzur 1996 or 2002), measurement of the maser radiation polarization can lead to an estimation of the magnetic field strength B and can reveal its spatial structure (via interferometric observations). In single dish observations, all of the maser components get smeared within the beam and only the mean value of B along the line of sight ($B_{//}$) can be derived. Only SiO masers are capable of tracing the magnetic field as close as $\sim$ 5 AU from the central star. But if SiO masers are to be used as a B-field tracer, we first need to give evidence that SiO masers are reliable B-field tracers. This requires more detailed theories than available today. Nevertheless, we tentatively derive in this work the field strength in the CSE inner layers of several evolved stars. 

Research on astronomical masers polarization is very active but is made difficult both by the lack of specific instrumental facilities and by the excitation and propagation of the masers themselves. Until now, numerous polarimetric observations of OH masers have been done, several of \water masers, but few of SiO maser emission. Few SiO polarimetric observations have been done with VLBI giving the very first images of the magnetic field  in some objects (e.g. Kemball \& Diamond 1997 in TX Cam). Most of the early SiO studies were done in linear polarization. The first complete SiO polarimetric observations were performed by Johnson \& Clark (1975), then by Troland \etal (1979); emission was found to be typically 15-30 \% linearly polarized and to exhibit no circular polarization. Barvainis, McIntosh \& Predmore (1987), and McIntosh \etal (1989) measured circular polarization of $1-9$ \% in several stars. Circular (0-4 \%) and linear (3.7-9.7 \%) polarizations were measured in VY CMa by McIntosh, Predmore \& Patel (1994). Later, Kemball \& Diamond (1997) made the first image of the magnetic field in the atmosphere of TX Cam, measuring a circular polarization level of 5 \% with some features showing polarization up to 30-40 \%.

It must be stressed that SiO, as H$_2$O, is a non-paramagnetic species. Zeeman splitting exists but the sublevels overlap; the effect is thus undetectable and hence only net polarization can be used to trace the magnetic field. The current status of our knowledge on the magnetic field strength can be summarized as follows: 
\begin{itemize}
  \item between 1000-10000 AU, $B_{//}\sim 5-20$ mG (OH masers, e.g. Kemball \& Diamond 1997, Szymczak \& Cohen 1997);
  \item at a few 100 AU from the star, $B_{//} \sim$ a few 100 mG (\water masers, e.g. Vlemmings, Diamond \& van Langevelde 2001, Vlemmings, van Langevelde \& Diamond 2005);
  \item at 5-10 AU, $B_{//} \sim 5-10$ G (SiO masers; Kemball \& Diamond 1997, in TX Cam). 
\end{itemize}

 The main purpose of this work is to measure and analyze the SiO maser polarization in terms of magnetic field strength in a representative sample of evolved stars. Our observations are presented in Section 2; they include simultaneous spectroscopic measurements of the 4 Stokes parameters. The results for individual stars are discussed in Section 3. In Section 4, we compare our data with predictions from existing SiO maser models and initiate a discussion on the validity of these models. Within the limitations of one of these models we derive the magnetic field strength and try to determine the role of the magnetic field. More broadly, a summary of the magnetic field topic in evolved stars is also given in Section 4. In Section 5 we give some concluding remarks.

\section{Observations}

An electromagnetic plane wave is defined by two components (horizontal and vertical):

\begin{equation}
\label{ }
e_H(z,t)=E_H\ e^{j(\omega t-kz-\delta)} 
\end{equation}
\begin{equation}
\label{ }
e_V(z,t)=E_V\ e^{j(\omega t-kz)}
\end{equation}

where $\delta$ is the phase difference between horizontal and vertical components.

Its energy flux is described by the 4 Stokes parameters:

\begin{equation}
\label{ }
I = <{E_H}^2> + <{E_V}^2>
\end{equation}
\begin{equation}
\label{ }
Q = <{E_H}^2> - <{E_V}^2>
\end{equation}
\begin{equation}
\label{ }
U = 2 <E_H E_V cos\  \delta>
\end{equation}
\begin{equation}
\label{ }
V = 2 <E_H E_V sin\  \delta>
\end{equation}

From these parameters, one deduces:

\begin{itemize}
  \item the circular polarization rate $p_C = V/I$
  \item the linear polarization rate $p_L = \sqrt{Q^2+U^2}/I$
  \item the polarization angle $\chi = \frac{\arctan (U/Q)}{2}$
\end{itemize}

The linear$/$circular polarization rate is sometimes called the linear$/$circular fractional polarization.

\begin{figure} [ht] 
  \begin{center} 
     \epsfxsize=7cm 
     \epsfbox{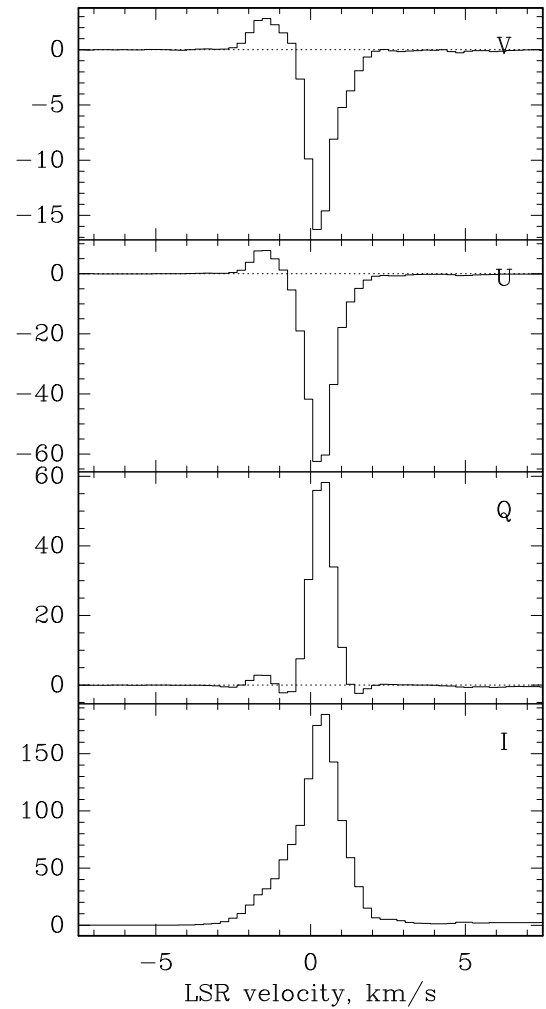} 
  \end{center} 
  \caption[]{V, U, Q and I Stokes parameters for R Leo. I, V, U and Q are given in Kelvins ($Ta^{\star}$; the conversion factor is 6 Jy$/$K).} 
  \label{} 
\end{figure} 

\scriptsize
\begin{table*} [htb]
 \caption{ \label{table} Stars observed in this work. The stellar type is derived from the literature as are the mass loss rates (e.g. Loup \etal 1993) and the period (e.g. AAVSO data). }
 {\begin{tabular}{l|c|c|c|c|c|c|c} \hline
{\bf Stars} &  RA & DEC & Type & $V_{LSR}$ & dM/dt & Period & Optical  \\
 & (J2000) & (J2000) & & [\kms] & [M$_{\odot}/$yr]& [days] &  phase \\ \hline
IRAS 18055-1433 & 18:08:23.20 & -14:32:43.0 & IR late-type & 180 & unknown & unknown &  \\
IRAS 18158-1527 & 18:18:41.50 & -15:26:25.0 & IR late-type & 20 & unknown & unknown &  \\ 
IRAS 18204-1344 & 18:23:17.90 & -13:42:46.0 & IR Supergiant (M8) & 45 &  4.2 $10^{-6}$& unknown & \\
W And & 02:17:32.96 & 44:18:17.8 & Mira (S6,1E-S9,2E$/$M4-M1) & -35 & 8.0 $10^{-7}$ & 395.9 & 0.62 \\
AU Aur & 04:54:15.00 & 49:54:00.3 & Mira (C6-7,3E(N0E)) & 8 & 1.1 $10^{-7}$ & 400 & 0.17   \\
NV Aur & 05:11:19.43 & 52:52:33.6 & Mira (M10) & 2 & 7.6 $10^{-6}$ & 635 &  \\
R Aur & 05:17:17.69 & 53:35:10.0 & Mira (M6.5E-M9.5E) & -3 & 9.8 $10^{-7}$ & 457.5 & 0.83 \\
RU Aur & 05:40:07.89 & 37:38:10.7 & SRb (M7E-M9E) & -35 & unknown & 466.4 & 0.25  \\
TX Cam & 05:00:51.15 & 56:10:54.0 & Mira (M8-M10) & 10 & 2.5 $10^{-6}$ & 557.4 & 0.70 \\
V Cam & 06:02:32.30 & 74:30:27.1 & Mira (M7E) & 8 & 1.6 $10^{-6}$ & 522.4 & 0.91 \\
R Cnc & 08:16:33.83 & 11:43:34.6 & Mira (M6E-M9E) & 14 & 6 $10^{-7}$ & 361.6 & 0.33  \\
W Cnc & 09:09:52.63 & 25:14:53.8 & Mira (M6.5E-M9E) & 38 & 3.1 $10^{-8}$ & 393.2 & 0.76 \\
VY CMa & 07:22:58.33 & -25:46:03.2 & Red Supergiant (M3-M4II) & 15 & $10^{-5}$ &
400 & 0.37  \\
S CMi & 07:32:43.08 & 08:19:05.3 & Mira (M6E-M8E) & 52 & 4.1 $10^{-8}$ & 332.9 & 0.35  \\
R Cas & 23:58:24.79 & 51:23:19.5 & Mira (M6E-M10E) & 27 & 1.1 $10^{-6}$ & 430.4 & 0.66 \\
S Cas & 01:19:41.97 & 72:36:39.3 & Mira (S3,4E-S5,8E) & -28 & 3.1 $10^{-6}$ & 612.4 & 0.98  \\
T Cas & 00:23:14.25 & 55:47:33.3 & Mira (M6E-M9.0E) & -7 & 5.1 $10^{-7}$ & 444.8 & 0.22  \\
T Cep & 21:09:31.85 & 68:29:27.6 & Mira (M5.5E-M8.8E) & -1 & 1.4 $10^{-7}$ & 388.1 & 0.94 \\
R Com & 12:04:15.20 & 18:46:56.7 & Mira (M5E-M8EP) & -3 & $10^{-7}$ & 362.8 & 0.43  \\
S CrB & 15:21:23.96 & 31:22:02.7 & Mira (M6E-M8E) & 3 & 5.8 $10^{-7}$ & 360.2 & 0.44  \\
R Crt & 11:00:33.87 & -18:19:29.6 & SRb (M7III) & 10 & 7.5 $10^{-7}$ & 160 & \\
$\chi$ Cyg & 19:50:33.94 & 32:54:50.6 & Mira (S6,2E-S10,4E) & 10 & 5.6 $10^{-7}$ & 408 & 0.37 \\
UX Cyg & 20:55:05.40 & 30:24:53 & irregular variable  (M4E-M6.5E) & 1 & 3.2 $10^{-6}$ & 565 & 0.77  \\
R Hya & 13:29:42.82 & -23:16:52.9 & Mira (M6E-M9E(TC)) & -8 & 1.4 $10^{-7}$ & 388.8 & 0.95  \\
W Hya & 13:49:02.03 & -28:22:03.0 & SRa (M7.5E-M9EP) & 42 & 8.1 $10^{-8}$ & 361 & 0.83 \\
X Hya & 09:35:30.26 & -14:41:28.5 & Mira (M7E-M8.5E) & 26 & 4.8 $10^{-8}$ & 301.1 & 0.0  \\
R Leo & 09:47:33.49 & 11:25:44.0 & Mira (M6E-M8IIIE-M9.5E) & 0 & $10^{-7}$ & 309.9 & 0.84 \\
W Leo & 10:53:34.44 & 13:42:54.4 & Mira (M5.5E-M7E) & 49 & unknown & 391.7 & 0.45  \\
R LMi & 09:45:34.28 & 34:30:42.8 & Mira (M6.5E-M9.0E) & 2 & 2.8 $10^{-7}$ & 372.2 & 0.56 \\
T Lep & 05:04:50.84 & -21:54:16.2 & Mira (M6E-M9E) & -29 & 7.3 $10^{-9}$ & 368.1 & 0.59  \\
RS Lib & 15:24:19.78 & -22:54:39.7 & Mira (M7E-M8.5E) & 7 & 1.8 $10^{-8}$ & 217.6 & 0.28 \\
Ap Lyn & 06:34:34.90 & 60:56:33.0 & Mira (M9) & -23 & 4.9 $10^{-6}$ & unknown &  \\
U Lyn & 06:40:46.49 & 59:52:01.6 & Mira (M7E-M9.5E) & -10 & unknown & 433.6 & 0.13  \\
GX Mon & 06:52:46.90 & 08:25:20.0 & Mira (M9) & -9 & 5.4 $10^{-6}$ & 527 &  \\
SY Mon & 06:37:31.28 & -01:23:43.6 & Mira (M6E-M9) & -57 & unknown & 422.2 & 0.18  \\
V Mon & 06:22:43.58 & -02:11:43.2 & Mira (M5E-M8E) & 5 & unknown & 341 & 0.0  \\
U Ori & 05:55:49.18 & 20:10:30.7 & Mira (M6E-M9.5E) & -38 & 2.8 $10^{-7}$ & 368.3 & 0.24 \\
RR Per & 02:28:28.73 & 51:16:21.1 & Mira (M6E-M7E) & 7 & unknown & 389.6 & 0.17 \\
S Per & 02:22.51.76 & 58:35:11.4 & SRc (M3IAE-M7) & -40 & 1.4 $10^{-6}$ & 822 & 0.58 \\
QX Pup & 07:42:16.83 & -14:42:52.1 & PN (M6) & 34 & 1.1 $10^{-4}$ & unknown & \\
Z Pup & 07:32:38.06 & -20:39:29.2 & Mira (M4E-M9E) & 4 & unknown & 508.6 & 0.74 \\
VX Sgr & 18:08:04.05 & -22:13:26.6 & Red Supergiant (M4EIA-M10EIA) & 6 & 5.5 $10^{-6}$ & 732 & 0.38  \\
AH Sco & 17:11:17.02 & -32:19:30.7 & SRc (M4E-M5IA-IAB) & -7 & $10^{-6}$ & 713.6 & 0.98  \\
RR Sco & 16:56:37.85 & -30:34:48.1 & Mira (M6II-IIIE-M9) & -28 & 1.1 $10^{-8}$ & 281.4 & 0.35  \\
R Ser & 15:50:41.74 & 15:08:01.4 & Mira (M5IIIE-M9E) & 28 & 2.6 $10^{-7}$ & 356.4 & 0.20  \\
S Ser & 15:21:39.53 & 14:18:53.1 & Mira (M5E-M6E) & 20 & $<2.2$ $10^{-7}$ & 371.8 & 0.76  \\
WX Ser & 15:27:47.30 & 19:33:48.0 & Mira (M8E) & 7 & 2.6 $10^{-6}$ & 425.1 & 0.30  \\
\hline
\end{tabular}}
\end{table*}
\normalsize

\addtocounter{table}{-1}
\scriptsize
\begin{table*} [htb]
 \caption{ \label{table} (-continued). Stars observed in this work. The stellar type is derived from the literature as are the mass loss rates (e.g. Loup \etal 1993) and the period (e.g. AAVSO data). }
 {\begin{tabular}{l|c|c|c|c|c|c|c} \hline
{\bf Stars} &  RA & DEC & Type & $V_{LSR}$ & dM/dt & Period & Optical  \\
 & (J2000) & (J2000) & & [\kms] & [M$_{\odot}/$yr]& [days] &  phase \\ \hline
IK Tau & 03:53:28.80 & 11:24:22.7 & Mira (M6E-M10E) & 35 & 4.4 $10^{-6}$ & 470 & 0.80 \\
R Tau & 04:28:18.00 & 10:09:44.8 & Mira (M5E-M9E) & 14 & 6.5 $10^{-8}$ & 320.9 & 0.68  \\
RX Tau & 04:38:14.57 & 08:20:09.4 & Mira (M6E-M7E) & -41 & $<5.8 10^{-8}$ & 331.8 & 0.13  \\
R Tri & 02:37:02.32 & 34:15:51.4 & Mira (M4IIIE-M8E) & 57 & 1.1 $10^{-7}$ & 266.9 & 0.0  \\
R UMi & 16:29:57.87 & 72:16:49.0 & SRb (M7IIIE) & -6 & unknown & 325.7 & 0.35 \\
S UMi & 15:29:34.66 & 78:38:00.2 & Mira (M6E-M9E) & -42 & unknown & 331 & 0.68 \\
R Vir & 12:38:29.95 & 06:59:19.0 & Mira (M3.5IIIE-M8.5E) & -26 & unknown & 145.6 & 0.12  \\
RS Vir & 14:27:16.39 & 04:40:41.1 & Mira (M6IIIE-M8E) & -12 & 3.8 $10^{-7}$ & 353.9 & 0.61  \\
RT Vir & 13:02:37.96 & 05:11:08.5 & SRb (M8III) & 18 & 7.4 $10^{-7}$ & 155 & 0.60 \\
S Vir & 13:33:00.11 & -07:11:41.0 & Mira (M6IIIE-M9.5E) & 12 & 4.1 $10^{-7}$ & 375.1 & 0.27  \\
\hline
\end{tabular}}
\end{table*}
\normalsize

We present here spectroscopic measurements of the 4 Stokes parameters (see Fig. 1). The observations were made with the IF polarimeter installed at the IRAM 30m telescope on Pico Veleta, Spain (Thum \etal 2003). Simultaneous measurements of I, U, Q, V allow us to calculate I,  $p_L$,  $p_C$ and $\chi$ for each velocity channel. The polarization angle calibration (i.e. the sign of Stokes U) was verified by observations of the Crab Nebula. Moreover, planets (polarization of planets is negligible at our frequency) have been used to check the instrumental polarization on the optical axis. 

The instrumental beam polarization is known to be stronger in Stokes Q and U than in Stokes V (known to be $\leq$ 2-3\%, see Thum \etal 2003, comparable to our sensitivity as stated elsewhere). If some detections from sources with weak p$_L$ are from a bad or uncertain pointing, they naturally induce a value of p$_C$ which is weaker than p$_L$. 

A strong instrumental polarization in Stokes V would be rather due to a bad phase tracking (the IF polarimeter works in a manner quite similar to that of an adding interferometer, and good phase tracking is essential). From several tests (Thum \etal 2003, Wiesemeyer, Thum \& Walmsley 2004), we know that polarization seen for weak SiO components with (Q,U,V)= (+ - -), (- - +) or (- + -) is instrumental polarization. We see that signature for only 3 objects (R Crt, R UMi and RT Vir). Some instrumental polarization may thus contaminate the observations of these objects.

All instrumental parameters were carefully calibrated through specific procedures described in  Thum \etal (2003). The error on  $p_{L,C}$ is $\leq 2-3$  \%. 

SiO (v=1, J=2-1) line observations at 86.243442 GHz were carried out towards 57 stars in August and November 1999 with the IRAM 30m radiotelescope. The pointing was regularly checked directly on the star itself (for the vast majority of objects). In order to obtain flat baselines, we used the wobbler switching mode. The system temperature of the SIS receiver ranged from 110 to 170 K. The front-ends were the facility receivers A100 and B100, and the back-end was the autocorrelator. The lines were observed with a spectral resolution of 0.3 \kms. The integration times were 4-10 minutes using the wobbler switching. The forward and main beam efficiencies were respectively 0.92 and 0.77 at 3 mm. (Additional SiO (v=1, J=5-4) line observations at 215.596 GHz were also performed in most stars studied here; results will be reported elsewhere.) 

Our source sample (see Table 1) consists of 43 Miras, 7 Semi-Regular stars (hereafter {\em SR}), 2 IR late-type stars, 1 irregular variable, 3 supergiants and 1 Planetary Nebula  (QX Pup) selected from our SiO maser master catalogue (Herpin \& Baudry, private communication). Coordinates and the main characteristics of the objects are given in Table 1. Nearly 60 \% of stars in this table have been observed with the HIPPARCOS satellite and have thus excellent optical positions; such positions have been adopted in our work.

\scriptsize
\begin{table} [htb]
 \caption{ \label{table} Derived parameters of the different components of the SiO
maser emission profile for each star. Only the well identified components are given (distinct peak or strong wing emission separated from the bulk emission).  Note that the polarization is fractional. The $\delta P$ is the rms derived from the $p_C$ plot.}
 {\begin{tabular}{l|c|c|c|c|c|c} \hline
{\bf Source} & v$_{LSR}$ & F$_{\nu}$ & p$_{c}$ & p$_{L}$ & $\delta$p & $\chi$  \\
 &  [\kms] & [Jy] &  & & & [$^{\circ}$] \\ \hline \hline
{\bf 18055-1433} & 180.6 & 1.02  & -0.10 & 0.11 & $  0.01$ & 50  \\
 & 182.7 & 1.7  &  -0.30 & 0.40 & $  0.01$ & 30  \\ \hline
{\bf 18158-1527} &  15.2 & 2.28  & 0.08 & 0.16 & 0.02 & 90   \\
 & 17.5 & 1.86 & -0.10 & 0.20 & 0.02 & 50   \\
 & 21.7 & 1.74  & 0.11 & 0.15 & 0.02 & 170   \\
 &  24.5 & 0.72  & 0.43 & 0.49 & 0.02 & 46   \\ \hline
{\bf 18204-1344} & 38.3 & 20.94  & 0.0 & 0.06 & $   0.01$ & 170   \\
 & 41.5 & 33.18 & 0.03 & 0.06 &  $   0.01$ &150   \\
 &  45.0 & 28.26 & 0.0 & 0.02 &  $   0.01$ &120  \\
 &  49.0 & 3.90& 0.04 & 0.08 &  $   0.01$ &110   \\
 & 53.7 & 8.34 & $\pm$0.07 & 0.06 &  $   0.01$ &120 \\  \hline
{\bf W And} & -38.0 & 4.98 & 0.08 & 0.20 &  0.01 & 70   \\
 & -35.9 & 16.32 & -0.01 & 0.02 & 0.01 & 30   \\
 &  -34.0 & 9.72 & -0.07 & 0.16 & 0.01 & 175   \\
 & -32.3 & 2.52 & 0.12 & 0.28 & 0.02 & 140  \\ \hline
{\bf AU Aur} & 3.4 & 3.72 & -0.09 & 0.17 &  $  0.01$ & 71   \\
 &  5.3 & 10.62 & 0.07 & 0.18 &  $  0.01$ & 165   \\
 &  7.9 & 27.12 & $\pm$0.02 & 0.09 &  $  0.01$ & 60  \\
 &  10.4 & 8.70 & 0.09 & 0.13 & $  0.01$ & 110 \\ \hline
{\bf NV Aur} &  -1.2 & 4.68 & -0.10 & 0.15 &  $  0.01$ & 140  \\
 &  1.7 & 17.94 & -0.06 & 0.13 &  $  0.01$ & 150  \\
 & 3.0 & 14.04 & -0.04 & 0.08 &  $  0.01$ & 170  \\ \hline
{\bf R Aur}  & -7.5 & 12.48 & 0.0 & 0.0 &  $  0.01$ & 60  \\
 & -5.6 & 31.56 & -0.09 & 0.27 &  $  0.01$ & 90  \\
 &  -3.7 & 36.00 & -0.13 & 0.34 &  $  0.01$ & 85 \\
 &  -1.2 & 38.58 & -0.07 & 0.16 &  $  0.01$ & 90 \\
 &  -0.1 & 13.14 & -0.12 & 0.27 &  $  0.01$ & 90  \\ \hline
{\bf RU Aur}  & -38.2 & 1.85 & 0.08 & 0.15 & 0.02 & 120  \\
 &  -34.9 & 1.05 & -0.12 & 0.40 & 0.02 & 85  \\
 &  -33.2 & 0.25 & -0.22 & 0.40 & 0.02 & 40  \\
 &  -27.1 & 0.2 & 0.0 & 0.35 & 0.02 & 175  \\ \hline
{\bf TX Cam}  & 5.0 & 11.52 &  -0.06 & 0.16 &  $  0.02$ & 130  \\
 &  6.1 & 13.68 & -0.03 & 0.08 &  $  0.02$ & 80  \\
 &  8.0 & 23.46 & 0.04 & 0.20 & 0.01 & 175  \\
 &  10.1 & 117.06 & -0.01 & 0.17 & 0.01 & 175  \\
 & 11.3 & 26.46 & 0.03 & 0.06 & 0.01 & 170  \\
 &  13.2 & 22.62 & -0.03 & 0.05 &  $  0.02$ & 50  \\
 &  14.9 & 12.78 & 0.02 & 0.18 &  $  0.02$ & 185  \\ \hline
{\bf V Cam}  & 3.8 & 3.12 & -0.04 & 0.22 & 0.01 & 0 \\
 &  7.6 & 12.72 & 0.0 & 0.04 & 0.01 & 160  \\
 &  9.0 & 9.24 & 0.02 & 0.05 & 0.01 & 0  \\ \hline
{\bf R Cnc}  & 9.2 & 2.70 & 0.12 & 0.26 &  $  0.01$ & 100  \\
 & 10.3 & 5.16 & -0.02 & 0.04 &  $  0.01$ & 160  \\
 & 13.7 & 56.46 & 0.02 & 0.17 &  $  0.01$ & 90  \\
  &  15.8 & 33.66 & -0.01 & 0.06 &  $  0.01$ & 180 \\ \hline
{\bf W Cnc}  & 33.9 & 21.18 &  0.13 & 0.32 & 0.01 & 40 \\
 &  41.9 & 2.82 & -0.03 & 0.12 & 0.01 & 150  \\ \hline
  \end{tabular}}
\end{table}
\normalsize

\addtocounter{table}{-1}
\scriptsize
\begin{table} [htb]
 \caption{ \label{table} (-continued). Derived parameters of the different components of the SiO
maser emission profile for each star. Only the well identified components are given (distinct peak or strong wing emission separated from the bulk emission).  Note that the polarization is fractional. The $\delta P$ is the rms derived from the $p_C$ plot.}
 {\begin{tabular}{l|c|c|c|c|c|c} \hline
{\bf Source} & v$_{LSR}$ & F$_{\nu}$ & p$_{c}$ & p$_{L}$ & $\delta$p & $\chi$  \\
 &  [\kms] & [Jy] &  & & & [$^{\circ}$] \\ \hline \hline
{\bf VY CMa} & 5.2 & 553.4 & -0.01 & 0.02 &  $  0.05$ & 125  \\
 & 8.5 & 973.3 & 0.0 & 0.01 &  $  0.05$ & 160 \\
 & 11.3 & 1086.7 & 0.04 & 0.06 &  $  0.05$ & 110  \\
 & 14.9 & 559.0 & -0.02 & 0.03 &  $  0.05$ & 80  \\
 & 17.6 & 553.0 & -0.03 & 0.08 &  $  0.05$ & 160  \\
 & 19.5 & 769.2 & 0.0 & 0.0 &  $  0.05$ & 160  \\
 & 22.7 & 1569.0 & -0.04 & 0.13 &  $  0.05$ & 0  \\
&  38.7 & 207.3 & -0.01 & 0.01 &  $  0.05$ & 150  \\ \hline
{\bf S CMi}  & 50.1 & 3.61 & 0.07 & 0.36 & 0.01 & 50  \\ 
 &  52.1 & 11.42 & 0.03 & 0.16 & 0.01 & 50  \\
 & 54.9 & 3.89 & 0.0 & 0.0 & 0.02 & 75  \\  \hline  
{\bf R Cas} & 25.4 & 600.0 & 0.06 & 0.20 &  $  0.01$ & 25  \\
 & 27.1 & 780.1 & 0.03 & 0.09 &  $  0.01$ & 70  \\
 &  28.4 & 419.6 & -0.02 & 0.08 &  $  0.01$ & 90 \\ \hline
{\bf S Cas} & -32.5 & 1.38 & 0.0 & 0.0 & 0.01 & 85 \\
 &  -30.0 & 11.52 & -0.32 & 0.52 & 0.01 & 110 \\
 &  -28.3 & 2.94 & 0.12 & 0.17 & 0.01 & 20 \\
 &  -27.2 & 4.98 & 0.05 & 0.05 & 0.01 & 20 \\ \hline 
{\bf T Cas} & -11.2 & 1.62 & 0.11 & 0.27 &  $  0.01$ & 110 \\
 & -8.7 & 11.70 & -0.02 & 0.18 &  $  0.01$ & 100 \\
 &  -5.1 & 5.82 & 0.08 & 0.12 &  $  0.01$ & 140  \\ \hline 
{\bf T Cep}  & -2.5 & 63.00 & $\pm$0.03 & 0.05 &  $  0.01$ & 100 \\
 &-0.5 & 117.42 & 0.04 & 0.16 &  $  0.01$ & 140  \\
 &  0.8 & 36.00& -0.04 & 0.05 &  $  0.01$ & 105\\  \hline 
{\bf R Com}  & -4.4 & 7.50 & -0.04 & 0.10 &  $  0.01$ & 80  \\
 &-3.3 & 8.82 & 0.07 & 0.27 &  $  0.01$ & 20 \\
  & -1.4 & 3.42 & 0.02 & 0.35 &  $  0.01$ & 50 \\ \hline 
{\bf S Crb}  & 0.8 & 41.40 & 0.07 & 0.17 &  $  0.01$ & 75  \\
  & 2.0 & 64.44 & 0.0 & 0.22 &  $  0.01$ & 70  \\
 & 4.9 & 28.62 & 0.0 & 0.35 &  $  0.01$ & 160  \\ \hline 
{\bf R Crt}  & 4.5 & 4.02 & -0.07 & 0.13 &  0.01 & 100 \\
 & 10.4 & 36.84 & 0.0 & 0.0 & 0.01 & 50 \\
& 16.4 & 4.74 & -0.08 & 0.15 & 0.01 & 120  \\ \hline 
{\bf $\chi$ Cyg}  & 7.1 & 50.16 & -0.19 & 0.42 &  $  0.01$ & 75  \\
 & 10.0 & 206.1 & 0.11 & 0.32 &  $  0.01$ & 140 \\
 & 14.6 & 36.72 & 0.0 & 0.0 &  $  0.01$ & 190  \\ \hline 
{\bf UX Cyg} & -1.3 & 9.12 & -0.04 & 0.08 &  $  0.01$ & 100  \\
 & -0.5 & 21.96 & 0.01 & 0.06 &  $  0.01$ & 110  \\
 & 0.8 & 44.22 & 0.02 & 0.06 &  $  0.01$ & 175   \\ \hline
{\bf R Hya} & -11.2 & 31.14 & 0.09 & 0.34 & 0.01 & 135  \\
 & -10.0 & 67.08 & 0.10 & 0.34 & 0.01 & 140  \\
& -8.9 & 182.82 & 0.02 & 0.06 & 0.01 & 150 \\
 & -6.6 & 146.22 & 0.05 & 0.25 & 0.01 & 130  \\ \hline 
{\bf W Hya} & 37.8 & 151.20 & 0.02 & 0.07 & 0.005 & 70  \\
& 41.4 & 934.38 & $\pm$0.02 & 0.03 & 0.005 & 150  \\
 & 44.8 & 202.02 & 0.06 & 0.15 & 0.005 & 100 \\ \hline
{\bf X Hya} & 21.8 & 4.26 & 0.01 & 0.70 & 0.01 & 80 \\
 & 23.6 & 3.18 & 0.03 & 0.25 & 0.01 & 60 \\
 & 28.9 & 23.28 & -0.01 & 0.08 & 0.01 & 170  \\ \hline 
{\bf R Leo}  & -1.4 & 195.30 & 0.10 & 0.30 &  $  0.01$ & 140 \\
 & 0.5 & 1110.6 & -0.09 & 0.36 &  $  0.01$ & 80  \\ \hline 
 \end{tabular}}
\end{table}
\normalsize

\addtocounter{table}{-1}
\scriptsize
\begin{table} [htb]
 \caption{ \label{table} (-continued). Derived parameters of the different components of the SiO
maser emission profile for each star. Only the well identified components are given (distinct peak or strong wing emission separated from the bulk emission).  Note that the polarization is fractional. The $\delta P$ is the rms derived from the $p_C$ plot.}
 {\begin{tabular}{l|c|c|c|c|c|c} \hline
{\bf Source} & v$_{LSR}$ & F$_{\nu}$ & p$_{c}$ & p$_{L}$ & $\delta$p & $\chi$  \\
 &  [\kms] & [Jy] &  & & & [$^{\circ}$] \\ \hline \hline
{\bf W Leo} & 42.7 & 2.04 & 0.0 & 0.18 & 0.05 & 170  \\
 & 44.6 & 2.76 & 0.0 & 0.10 & 0.05 & 100  \\
 & 47.9 & 3.30 & 0.0 & 0.20 & 0.05 & 170  \\
& 54.2 & 4.62 & 0.08 & 0.42 & 0.02 & 35 \\ \hline
{\bf R LMi}  & 0.3 & 61.98 & 0.0 & 0.02 & 0.01 & 25  \\
 & 2.2 & 52.80& 0.12 & 0.36 & 0.01 & 0  \\
 & 3.5 & 34. 62 & 0.06 & 0.20 & 0.01 & 15 \\
 & 4.5 & 14.16 & 0.04 & 0.08 & 0.01 & 20 \\
 & 6.2 & 10.56 & 0.05 & 0.20 & 0.01 & 80  \\  \hline  
{\bf T Lep} & -32.0 & 15.48 & -0.11 & 0.34 &  $  0.01$ & 10  \\
& -30.1 & 10.74 & -0.06 & 0.20 &  $  0.01$ & 20 \\
 & -27.1 & 42.48 & 0.03 & 0.10 &  $  0.01$ & 95  \\ 
 & -25.4 & 7.80 & 0.10 & 0.14 &  $  0.01$ & 75 \\ \hline
{\bf RS Lib} & 3.2 & 7.62 & 0.01 & 0.08 & 0.01 & 150  \\
 & 6.5 & 35.94 & -0.10 & 0.15 & 0.01 & 150 \\
 & 9.1 & 12.24 & -0.06 & 0.20 & 0.01 & 150 \\ \hline
{\bf Ap Lyn} & -25.1 & 9.90 & 0.02 & 0.37 & $  0.01$ & 185  \\
 & -23.0 & 36.36 & 0.12 & 0.65 &  $  0.01$ & 165 \\
& -20.9 & 9.24 & 0.03 & 0.20 &  $  0.01$ & 150 \\ \hline
{\bf U Lyn}  & -14.0 & 12.00 & -0.02 & 0.04 & 0.01 & 45  \\
  & -11.7 & 30.78 & 0.0 & 0.08 & 0.01 & 40 \\
 & -4.5 & 7.19 & -0.13 & 0.24 & 0.01 & 80  \\ \hline 
{\bf GX Mon} & -10.4 & 29.64 & 0.12 & 0.38 &  $  0.01$ & 100  \\
 & -8.5 & 21.60 & -0.03 & 0.08 &  $  0.01$ & 20 \\
  & -6.4 & 10.38 & 0.07 & 0.20 &  $  0.01$ & 120 \\
   & -4.9 &  7.92 & -0.02 & 0.07 &  $  0.01$ & 160  \\
   & -3.0 & 4.98 & 0.01 & 0.02 &  $  0.01$ & 55 \\ \hline
{\bf SY Mon}   & -59.6 & 3.01 & 0.12 & 0.58 & 0.02 & 75 \\
   & -56.0 & 2.52 & -0.05 & 0.15 & 0.02 & 30  \\ \hline 
{\bf V Mon}   & 2.0 & 8.58 & 0.05 & 0.10 & 0.02 & 170 \\
   & 4.8 & 3.12 & 0.05 & 0.14 & 0.02 & 80 \\ \hline
{\bf U Ori}   & -42.4 & 12.42 & 0.05 & 0.15 & 0.01 & 80  \\
   & -39.9 & 46.26 & -0.04 & 0.16 & 0.01 & 50  \\
   & -37.8 & 13.08 & 0.02 & 0.06 & 0.01 & 145 \\
   & -34.2 & 32.40 & 0.13 & 0.40 & 0.01 & 170 \\ \hline
{\bf RR Per}   & 5.1 & 10.81 & 0.03 & 0.14 & 0.01 & 165  \\
   & 7.5 & 29.46 & 0.11 & 0.43 & 0.01 & 175  \\
   & 8.6 & 27.78 & 0.03 & 0.30 & 0.01 & 185 \\
   & 11.1 & 13.02 & 0.10 & 0.34 & 0.01 & 180  \\ \hline
{\bf S Per}   & -47.4 & 7.02 & -0.03 & 0.03 & 0.02 & 75  \\
   & -44.9 & 31.62 & 0.15 & 0.04 & 0.05 & 190 \\
   & -43.3 & 35.70 & 0.01 & 0.04 & 0.05 & 190  \\
   & -39.2 & 71.16& 0.0 & 0.03 & 0.05 & 150  \\
   & -36.0 & 37.79 & $\pm$0.03 & 0.06 & 0.02 & 80 \\ \hline 
 {\bf QX Pup}   & 27.3 & 11.22 & 0.02 & 0.15 &  $  0.01$ & 120  \\
   & 29.4 & 6.84 & 0.0 & 0.03 &  $  0.01$ & 70  \\
   & 35.3 & 9.12 &  0.17 & 0.41 &  $  0.01$ & 120 \\  
   & 41.6 & 2.70 & 0.04 & 0.07 &  $  0.01$ & 10  \\ \hline
{\bf Z Pup}   & 3.9 & 40.98 & 0.07 & 0.23 & 0.01 & 140  \\ \hline
{\bf VX Sgr}   & -6.1 & 25.98 & 0.0 & 0.0 &  $  0.005$ & 140  \\
   & 0.9 & 89.16 & 0.01 & 0.02 &  $  0.005$ & 130  \\
   & 3.4 & 122.39 & -0.02 & 0.05 &  $  0.005$ & 10  \\
   & 5.7 & 178.38 & -0.01 & 0.04 &  $  0.005$ & 10 \\
   & 10.1 & 114.61 & -0.03 & 0.10 &  $  0.005$ & 145  \\
   & 14.1 & 53.34 & 0.01 & 0.01 &  $  0.005$ & 145 \\
   & 16.2 & 80.04 & 0.01 & 0.02 &  $  0.005$ & 160  \\ \hline
\end{tabular}}
\end{table}
\normalsize

\addtocounter{table}{-1}
\scriptsize
\begin{table} [htb]
 \caption{ \label{table} (-continued). Derived parameters of the different components of the SiO
maser emission profile for each star. Only the well identified components are given (distinct peak or strong wing emission separated from the bulk emission).  Note that the polarization is fractional. The $\delta P$ is the rms derived from the $p_C$ plot.}
 {\begin{tabular}{l|c|c|c|c|c|c} \hline
{\bf Source} & v$_{LSR}$ & F$_{\nu}$ & p$_{c}$ & p$_{L}$ & $\delta$p & $\chi$  \\
 &  [\kms] & [Jy] &  & & & [$^{\circ}$] \\ \hline \hline
{\bf AH Sco}  & -11.4 &  26.69 & 0.05 & 0.06 &  $   0.005$ &120  \\
   & -10.0 & 48.31 & 0.01 & 0.02 &  $   0.005$ &120  \\
   & -7.2 & 78.78 & 0.0 & 0.0 &  $   0.005$ &150 \\
   & -4.7 & 54.74 & 0.0 & 0.02 &  $   0.005$ &20  \\
   & -2.8 & 32.99 & -0.02 & 0.05 &  $   0.005$ &170  \\ \hline
{\bf RR Sco}   & -33.9 & 4.02 & 0.10 & 0.14 & 0.02 & 65 \\
   & -30.2 & 9.48 & 0.0 & $\pm$0.07 & 0.02 & 85 \\
   & -26.5 & 11.16 & 0.04 & 0.10 & 0.02 & 90 \\
   & -23.2 & 2.82 & 0.06 & 0.16 & 0.02 & 55  \\ \hline 
{\bf R Ser} & 27.8 & 21.61 & -0.07 & 0.30 &  $  0.01$ & 40  \\ \hline 
{\bf S Ser}   & 18.7 & 11.22 & 0.04 & 0.09 &  $  0.01$ & 180  \\
   & 20.9 & 16.92 & 0.0 & 0.06 &  $  0.01$ & 100  \\ \hline
{\bf WX Ser}   & 4.1 & 20.34 & -0.05 & 0.15 &  $  0.01$ & 140  \\
   & 6.6 & 12.18 & 0.04 & 0.15 &  $  0.01$ & 80 \\
   & 9.8 & 8.41 & -0.12 & 0.30 &  $  0.01$ & 140  \\ \hline
{\bf IK Tau}   & 31.6 & 46.74 & 0.04 & 0.14 &  0.01 & 180  \\
  & 34.7 & 288.24 & 0.05 & 0.16 & 0.01 & 170 \\
   & 36.8 & 53.04 & 0.07 & 0.18 & 0.01 & 140 \\
   & 39.6 & 27.84 & 0.13 & 0.36 & 0.01 & 125 \\ \hline
{\bf R Tau}   & 9.5 & 3.78 & -0.10 & 0.15 &  $  0.01$ & 60  \\
   & 11.8 & 19.51 & 0.06 & 0.13 &  $  0.01$ & 150  \\
   & 13.1 & 10.26 & 0.07 & 0.19 &  $  0.01$ & 140 \\
   & 14.7 & 7.68 & -0.09 & 0.31 &  $  0.01$ & 55  \\
   & 16.4 & 6.24& -0.08 & 0.19 &  $  0.01$ & 75 \\
   & 19.0 & 3.24 & -0.20 & 0.44 &  $  0.01$ & 55 \\ \hline 
{\bf RX Tau}   & -44.0 & 3.36 & -0.06 & 0.10 & 0.02 & 120  \\
   & -41.0 & 10.98 & -0.01 & 0.01 & 0.02 & 100  \\
   & -38.9 & 2.94 & 0.04 & 0.15 & 0.02 & 170  \\ \hline 
{\bf R Tri}   & 56.7 & 13.62 & -0.05 & 0.24 &  $  0.01$ & 105 \\ \hline 
{\bf R UMi}   & -6.1 & 1.98 & 0.0 & 0.25 & 0.05 & 110  \\ \hline
{\bf S UMi}  & -44.4 & 21.54 & 0.0 & 0.0 & 0.005 & 30  \\
   & -43.3 & 30.66 & 0.0 & 0.01 & 0.005 & 20  \\
   & -40.8 & 24.24 & -0.02 & 0.10 & 0.005 & 10 \\
   & -40.0 & 27.66 & 0.02 & 0.07 & 0.005 & 10 \\ \hline 
{\bf R Vir}    & -26.8 & 0.84 & 0.15 & 0.30 & 0.05 & 170  \\
   & -25.7 & 2.16 & -0.10 & 0.15 & 0.02 & 10  \\ \hline
{\bf RS Vir}  & -15.7& 2.64 & -0.11 & 0.18 & 0.01 & 0 \\
   & -13.3 & 10.62 & -0.08 & 0.32 & 0.01 & 150  \\
   & -12.5 & 8.46 & 0.0 & 0.20 & 0.01 & 0  \\
   & -9.1 & 7.08 & -0.06 & 0.52 & 0.01 & 160 \\ \hline
{\bf RT Vir}   & 9.1 & 1.20 & 0.0 & 0.02 &  $  0.02$ & 70   \\
   & 13.8 & 2.22 & 0.12 & 0.19 &  $  0.02$ & 20  \\
   & 15.7 & 3.12 & 0.12 & 0.20 &  $  0.02$ & 140  \\
   & 18.0 & 7.98 & 0.0 & 0.06 &  $  0.02$ & 130  \\
   & 20.7 & 3.48 & -0.10 & 0.12 &  $  0.02$ & 120  \\
   & 22.6 & 3.03 & -0.06 & 0.11 &  $  0.02$ & 60 \\
   & 27.7 & 2.34 & -0.04 & 0.10 &  $  0.02$ & 40 \\ \hline 
{\bf S Vir}   & 10.3 & 13.62 & -0.04 & 0.27 &  $  0.01$ & 180  \\
  & 12.2 & 7.79 & 0.03 & 0.06 &  $  0.01$ & 105  \\
   & 13.7 & 22.52 & -0.05 & 0.19 &  $  0.01$ & 130  \\ \hline 
\end{tabular}}
\end{table}
\normalsize

\section{Results: Polarization Study}

\subsection{Individual results}

Values of the polarization level presented here (see Table 2) are those measured for the different components within the SiO
maser emission profile for each star. Examples are given in Fig.2 for a few stars. The complete Figure 2 with all the observations is available in electronic form at http://www.edpscience.org. Only the well identified components are considered (distinct peaks or strong wing emission well separated from the bulk emission, according to the noise).  Some interesting cases  are briefly presented below.

 Some profiles show isolated emission red$/$blue-shifted from the main emission which are more strongly circularly polarized (e.g. IRAS 18204-1344). These peculiar characteristics imply a different spatial origin for the main and higher/lower velocity components. Sometimes the circular polarization is regularly varying across the profile (e.g. R Leo), but sometimes not.

In T Lep, the SiO emission shows two peaks linked by a plateau; the circular polarization is linearly varying across the profile from -11 to 10 \% (see Fig. 2). Several objects (e.g. S UMi, IK Tau) show the same $p_C$ pattern. 

The IR late-type source  IRAS18158-1527 exhibits a complex profile with several well defined components, each of them differently polarized indicating a complex maser structure with probably different maser spots contributing to the whole emission. The red wing emission is highly polarized (43 \%). Such a complex multi-component maser line profile and "semi-circle", convex, $p_C$ pattern appear to be characteristic of SR objects (see other similar objects in our sample and R Crt in Fig. 2). Nevertheless, the circular polarization pattern observed in the Mira star U Lyn is a convex profile as encountered in SR objects. 

One of the most studied Mira star is R Leo. The profile is made of a strong emission with a blue broad line wing. Main and linewing emissions are strongly polarized (respectively negative and positive $p_C \sim 9-10$ \%). R Leo is a very well studied object exhibiting a bipolar jet throughout its envelope. The clear symmetry observed between the positive and negative circular polarization patterns in the main and wing line emissions suggests that the maser emission comes from the jet lobes. We note that the Mira star RS Lib exhibits an emission and polarization pattern similar to that observed in R Leo. R Leo and RS Lib may have the same spatial structure. 

\begin{figure*} [ht] 
  \begin{center} 
     \epsfxsize=11cm 
     \epsfbox{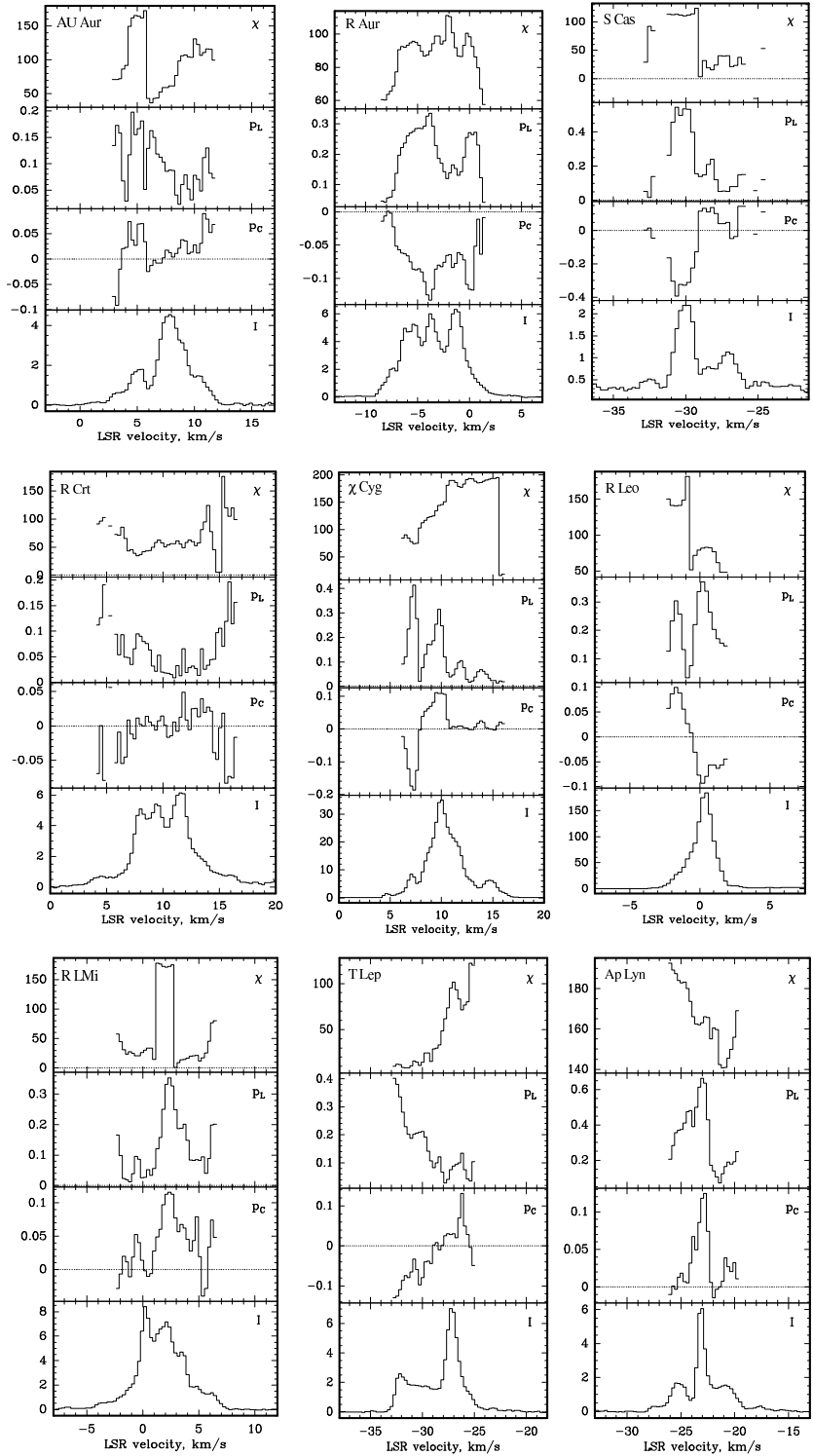} 
  \end{center} 
  \caption []{Position angle of polarization ($\chi$) in degrees, linear ($p_L$) and circular ($p_C$) polarization levels and intensity ($I= Ta^{\star}$ in Kelvin; the conversion factor is 6 Jy$/$K) for the SiO emission observed in several stars. } 
  \label{} 
\end{figure*} 

  \addtocounter{figure}{-1}
\begin{figure*} [ht] 
  \begin{center} 
    \epsfxsize=11cm 
     \epsfbox{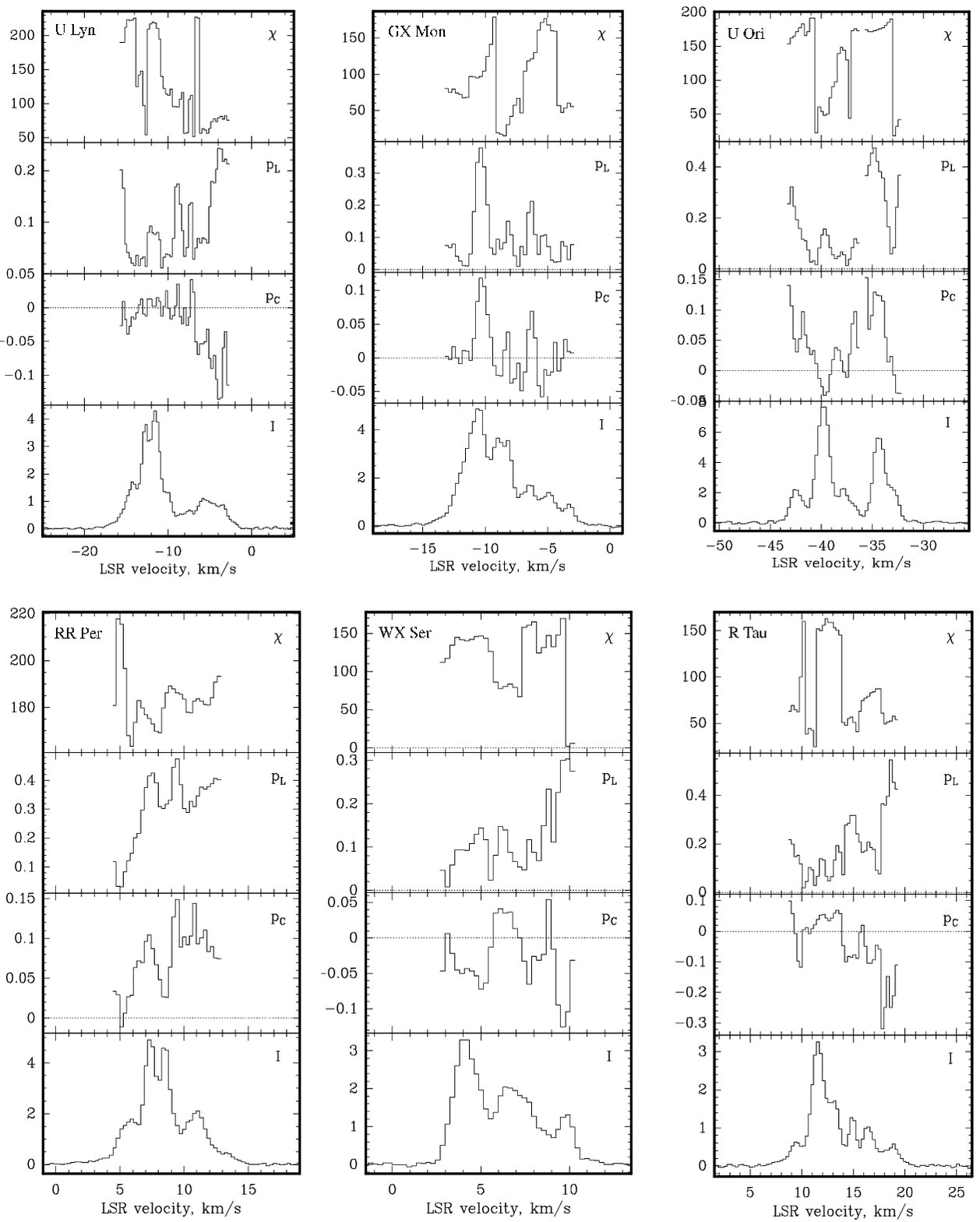} 
  \end{center} 
  \caption []{(-continued). Position angle of polarization ($\chi$) in degrees, linear ($p_L$) and circular ($p_C$) polarization levels and intensity ($I= Ta^{\star}$ in Kelvin) for the SiO emission observed in several stars. } 
  \label{} 
\end{figure*}

\subsection{Analysis}

 The circular polarization level in several of our objects has already been measured by Kemball \& Diamond (1997) or Barvainis, McIntosh \& Predmore (1987). For TX Cam and W Hya, our results are consistent with previous observations:
\begin{itemize}
  \item in TX Cam, the bulk of the emission is weakly circularly polarized while its wings show  $p_C \sim 3-6 \%$ in good agreement with the VLBI observations of Kemball \& Diamond in 1997 who derived an average value of $p_C \sim 3-5 \%$;
  \item in the Semi-Regular object W Hya, the central emission is weakly polarized ($\pm$ 2 \%), while the wings and secondary peaks show  $p_C$= 2-6 \%), which is consistent with the 5 \% of Barvainis, McIntosh \& Predmore (1987).
 \end{itemize} 
On the contrary, the circular polarization level we derive in VY CMa, R Cas, R Leo, and VX Sgr is different from  levels measured by Barvainis, McIntosh \& Predmore (1987), respectively 1-4, 2-6, 9-10 and less than 3 \% while they found respectively 6.5, 1.5, 2.4 and 8.7 \%. 
This difference is significant and may be due to variability over the fifteen intervening years. Indeed time variability of the polarization remains an open question in the field. Glenn \etal (2003) have shown that the individual maser feature lifetime ranges from a few months or less to more than 2 years, i.e. the characteristic time over which the $Q$ and $U$ spectral features persist. The average linear polarization is 23 \% in Glenn \etal sample with a typical dispersion of 7\%. Cotton et al. (2004) have comparable epoch spacing and do not conclude on the variability. Our observations were repeated at intervals of a few months (August and November 1999) and the polarization tends to remain stable between the two epochs.

We emphasize that we cannot spatially distinguish with a single dish radiotelescope between the different maser spots producing the SiO profile (various masers spots contribute in the various features observed at a given velocity). The whole SiO maser emission region, hence all the maser cells, lie within the 29 arcseconds of the 30m  (but not necessarily with a uniform distribution) while the SiO emission covers less than 40 milliarseconds in TX Cam (Kemball \& Diamond 1997) and thus everything is beam averaged. This means that any conclusion on the geometry of the objects observed here would be much uncertain. Only global trends or global geometry can be discussed. One of the consequences of this spatial resolution problem is that if the polarization vectors are distributed isotropically around the object, the average polarization level that we measure is zero, even if the maser emission produced in each SiO cell is well polarized.

A global analysis of our data in Table 2 shows the following. We find that $p_L$ varies between 0 and 70 \%, and  $p_C$ between 0 and $\pm$43 \%. Hence, polarization vectors are not distributed isotropically. Emission from Mira-type objects clearly tends to have a relatively high linear  ( ${<p_L>}_{Mira} \simeq 30$\%, ${<p_L>}_{SR} \simeq 11$\%) and circular polarization  (${<p_C>}_{Mira} \simeq 9$\%, ${<p_C>}_{SR} \simeq 5$\%). Note that the emission from the PN QX Pup is highly polarized, and, on the contrary, maser emission from supergiants shows very weak polarization ($<p_L>=5$\%, $<p_C>=2$\%), with the exception of one maser component in S Per. Moreover, all observations show that the polarization level varies across the maser line profile (see Fig. 2), i.e. the different spectral components of the maser emission producing the profile are coming from different localizations in the SiO shell and have different polarization levels. The highest polarization level for one object can be encountered either in the main peak, or in the other components.

 Semi-Regular objects (RU Aur, R Crt, W Hya, S Per, AH Sco, R UMi, RT Vir) have a common circular polarization pattern with the central main emission unpolarized and other peak emission or wings being strongly polarized:  a characteristic "semi-circle" (i.e. convex shape) pattern for $p_C$ is observed (see R Crt in Fig. 2). The infrared late-type star IRAS18158-1527 exhibits a similar pattern, thus suggesting that this star is a semi-regular.

A group of objects (W And, NV Aur, T Cas, R Com, T Lep, IK Tau, S Ser, S UMi) shows approximately the same $p_C$ pattern  (see T Lep in Fig. 2); the circular polarization varies linearly across the line profile from a positive value to a negative one (or the contrary). The only common spectral characteristic of the SiO emission from these stars is the presence of an plateau-like emission on top of which the narrow emission peaks are located. 
 

\begin{figure*}
\begin{center}
\includegraphics[width=4in, angle=270]{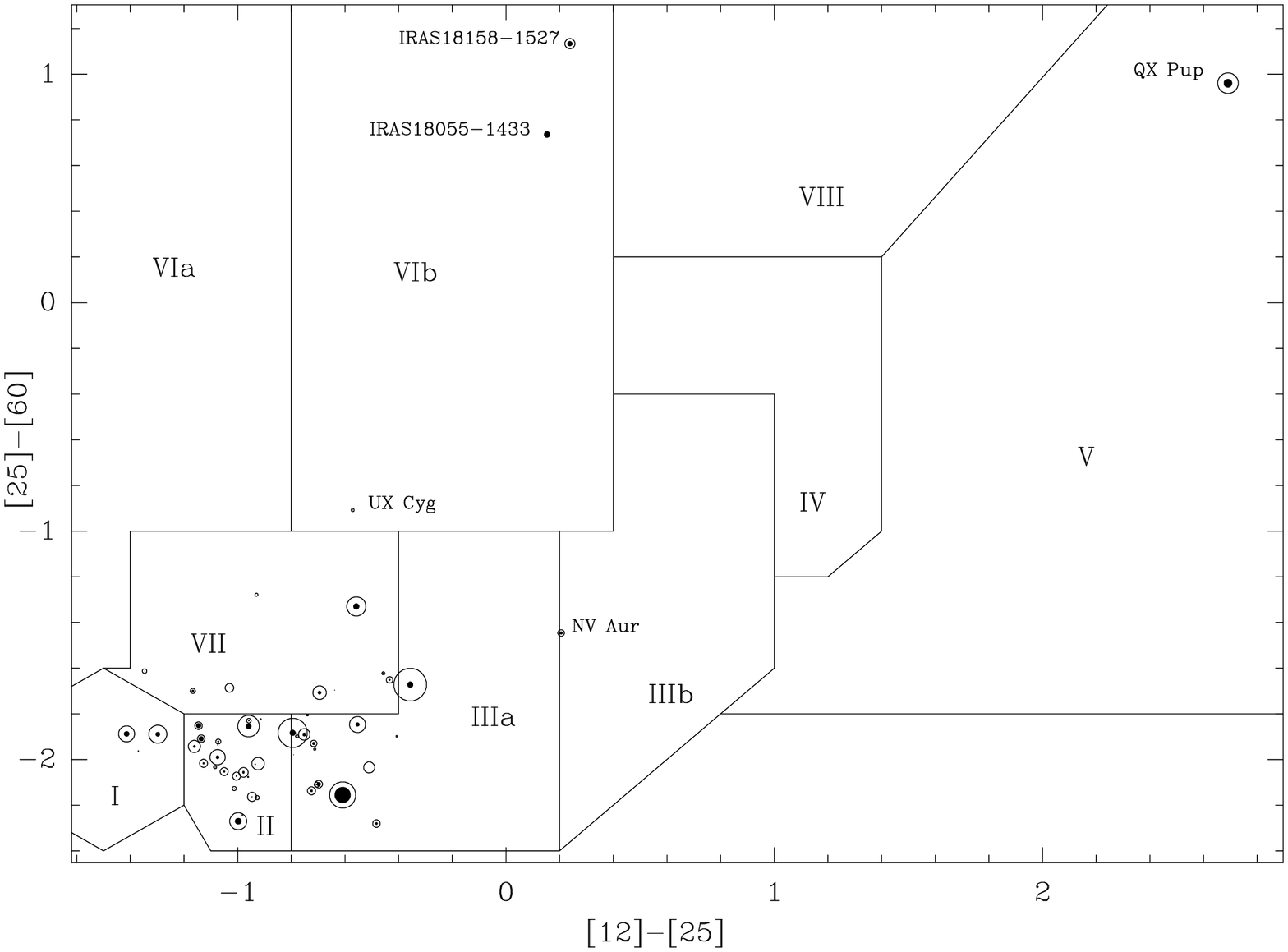}
\caption{Two-color diagram for our sample. The size of the empty and filled circles for each star are proportional to respectively the linear and absolute circular polarization level measured for the main peak of SiO emission. }
\label{ }
\end{center}
\end{figure*}

\section{Discussion}

In this section, we first discuss our source sample in the frame of the 2-color diagram. Then, we briefly summarize the existing SiO maser polarization theories. Finally, we discuss our data set in this context and estimate the stellar magnetic field strength.

\subsection{2-color diagram}

Stars of our sample can be plotted in a [12]-[25], [25]-[60] color-color diagram (van der Veen \& Habing 1988; [12], [25] and [60]  stand respectively for 12, 25  and 60 microns IRAS-fluxes). This diagram is partitioned into several regions (see Fig. 3) defined by van der Veen \& Habing as follows: 
Region I, oxygen-rich non variable stars without circumstellar shells; 
Region II, variable stars with young O-rich circumstellar shells; Region IIIa, 
variable stars with more evolved O-rich circumstellar shells; Region IIIb, 
variable stars with thick O-rich circumstellar shells; Region IV, variable stars 
with very thick O-rich circumstellar shells; Region V, Planetary Nebulae and non-variable stars with very cool envelopes; Region VIa, non variable stars 
with relatively cold dust at large distance; Region VIb, variable stars with
relatively hot dust close to the star and relatively cold dust at large
distance; Region VII, variable stars with more evolved C-rich circumstellar
shells.  

On Fig. 3 are represented the linear and circular polarization level for the main SiO emission component from each star. Most of the objects in our sample fall in regions II and IIIa and do not show particular characteristics, except for S Cas (an S-type star) where the circular polarization is high. Mira-type stars are in regions I, II, IIIa and VII. IR late-type objects are in VIb and VII. The SRa semi-regular variable W Hya is in I, while the SRb stars lie in IIIa (RU Aur, RT Vir, R Crt), II (R UMi), and Src in IIIa (S Per) and VII (AH Sco). The Red Supergiants, VY CMa and VX Sgr, and the IR supergiant IRAS 18204-1344 lie in VII. Objects in Region VII do not exhibit strong polarization compared to other objects, perhaps because of their more C-rich circumstellar shells (e.g. AU Aur) or because of the presence of hot dust close to the star implying less SiO abundance and thus weaker emission, making the polarization measurement less significant. The presence of hot dust may also influence the pumping of the SiO molecules and thus the polarization level; the optically thick, hence isotropic, radiation field of hot dust can assist the collisional pumping. This could apply to UX Cyg (an irregular variable), IRAS 18055-1433 and IRAS 18158-1527 in region VIb. QX Pup in region V is a PN and exhibits strong polarized emission. Note that IRAS 18055-1433 and IRAS 18158-1527 show very strong circular polarization in their line wings. We may conjecture here that wing emission comes from more outer layers than those where the main line is excited (Herpin \etal 1998); as a consequence, the SiO cells giving rise to wing emission are less influenced by the presence of hot dust (hot dust preferably lies in the inner layers).

\subsection{The SiO maser polarization theory}

 Since SiO is non-paramagnetic, the Zeeman splitting $g\nu_B$ ($g$ is the LandŽ factor) is much less than the Doppler width. Moreover, the degree of saturation is the ratio of the rate $R$ for stimulated emission to the loss rate $\Gamma$ (usually $\Gamma$ is approximated by the inverse radiation lifetime for a vibrational transition, $\Gamma \simeq 5$ s$^{-1}$ for SiO masers, Wiebe \& Watson 1998). Hence, if $R\geq \Gamma$, the maser is saturated. In fact, in the Orion case Plambeck \etal (2003) show that, despite the radiation beam angle is unknown, the 86 GHz SiO maser is saturated. The maser is saturated if the angle averaged intensity $J= \frac {I \Omega_b} {4\pi}$ ($\Omega_b$ is the beaming solid angle) is larger than the saturation intensity $J_S$; $J_S$ is a theoretical quantity. The saturation depends on the angle into which the radiation is beamed, but this angle is unknown (Watson \& Wyld 2001), thus $J$ cannot be directly measured (even if $I$ is measurable when the maser is resolved, the beaming angle $\Omega$ is not an observable).

For more than one decade, two schools have come up against each other to explain SiO maser emission. SiO polarization theory is described in: (i) Watson (e.g. Watson \& Wyld 2001, Wiebe \& Watson 1998, Nedoluha \& Watson 1994); (ii)  Elitzur (2002, 1998, 1996, 1994).

The main difference between the two approaches rests in the pumping mechanisms. While anisotropic pumping associated with a weak field produces high $p_L$ and quite significant $p_C$ in Watson's model, a strong magnetic field is necessary with the more classical pumping mechanisms used in Elitzur's model. Details about both models can be found in the Elitzur's review (2002).

We may summarize the main characteristics of Watson's model as follows:
\begin{itemize}
	\item non-Zeeman effect;
  \item anisotropic pumping;
  \item no direct relation between $p_{C}$ and $B$. An estimation of B can only be derived through complete calculation of the radiative transfer. Nevertheless, when maser saturation is not important, the "thermal" spectral line equation $\frac {V}{\delta I/\delta v}=\alpha B \cos \theta$ is applicable (Fiebig \& G\"usten 1989); $I$ is the intensity with respect to Doppler velocity $v$, $\theta$ is the angle between $B$ and the line of sight, $\alpha$ is a constant;
    \item the Zeeman splitting parameter $g\Omega$ (in frequency units $g\Omega = 1.5 B[mG] s^{-1}$ for SiO masers) is $\simeq R$;
    \item  saturated or unsaturated maser (saturated maser increases $p_C$);
    \item  linear correlation between $p_L$ and $p_C$, and high $p_L$ is needed; 
    \item  intensity dependent circular polarization; 
    \item  $B$ of a few 10 mG varying as $r^{-2,3}$ throughout the envelope. 
 \end{itemize}

In contrast with Watson's work, Elitzur's model is based on the Zeeman effect and the exponential maser growth in the unsaturated phase; the polarization characteristics are preserved as the radiation is amplified into the saturated regime. This model was improved several times (Elitzur 1994, 1996, 1998)  and takes into account the anisotropic pumping. The magnetic field generates circular polarization  and the main pumping mechanism for the SiO maser is a "classical" radiative-collisional process. For saturated masers, a direct relation between $p_C$ and $B$ is obtained from simple calculations. The ratio $x_B$ of the Zeeman splitting $\Delta \nu_B$ to the Doppler linewidth $\Delta \nu_D$, can be determined (Elitzur 1996) from $v_{peak}$, the ratio of the Stokes parameter $V/I$ at a given peak feature:
\begin{equation}
\label{ }
x_B=\frac {3\sqrt{2}}{16} ~ v_{peak} \cos \theta
\end{equation}
Following Barvainis, Mc Intosh  \& Predmore (1987) and Elitzur (1996) we arbitrarily take $\theta \simeq 45^{\circ}$
\begin{equation}
\label{ }
\Rightarrow x_B= \frac{3}{16} ~ v_{peak}
\end{equation}
\begin{equation}
\label{ }
\Rightarrow x_B=  0.1875 ~10^{-2} ~m_C
\end{equation}
where $m_C$ is the polarization fraction in percentage (i.e. 100 $p_C$). 
Moreover: 
\begin{equation}
\label{ }
x_B= 14 g \lambda \frac {B}{\Delta v_D}
\end{equation}
where $g$ is the Land\'e factor with respect to the Bohr magneton, $\lambda$ the transition wavelength in cm, $B$ the field in Gauss and $\Delta v_D$ the Doppler width in km$s^{-1}$. Thus,
\begin{equation}
\label{ }
B= 0.1875 ~10^{-2} ~\frac {\Delta v_D}{14 g \lambda} ~m_C
\end{equation}
With $g\simeq 10^{-3}$, $\Delta v_D =1$ \kms and $\lambda=0.2877$ cm, we derive:
\begin{equation}
\label{ }
B\simeq 0.46 ~m_C
\end{equation}

This model predicts that there is no correlation between $p_C$ and $p_L $ (see also Watson \& Wyld 2001); such a mechanism leads to an inferred magnetic field of a few Gauss to 10 Gauss for the SiO maser zone, the field hence varying as $r^{-1,2}$ across the envelope. 

\subsection{Theory against present observations}

The relevance of the two different polarization theories can be assessed via a few observational checks:
\begin{itemize}
  \item dependence of circular polarization on intensity;
  \item linear correlation between $p_C$ and $p_L$;
  \item Zeeman effect, i.e. spectral shape of the Stokes parameter $V$;
  \item coherence of B strength values inferred from OH and \water observations.
\end{itemize}
As we do not resolve the maser emission into individual spatial components, our current data set is biased by the beam averaging. Therefore, some effects cannot be tested with our data. In the Zeeman effect case, the spectral shape of the Stokes parameter $V$ must follow an antisymmetric {\em S} curve with sharp reversal at line center (Elitzur 1996). Unfortunately the doppler width is less than the resolution of the observations and we cannot conclude. 
 
\begin{figure}
\begin{center}
\includegraphics[width=2.4in, angle=270]{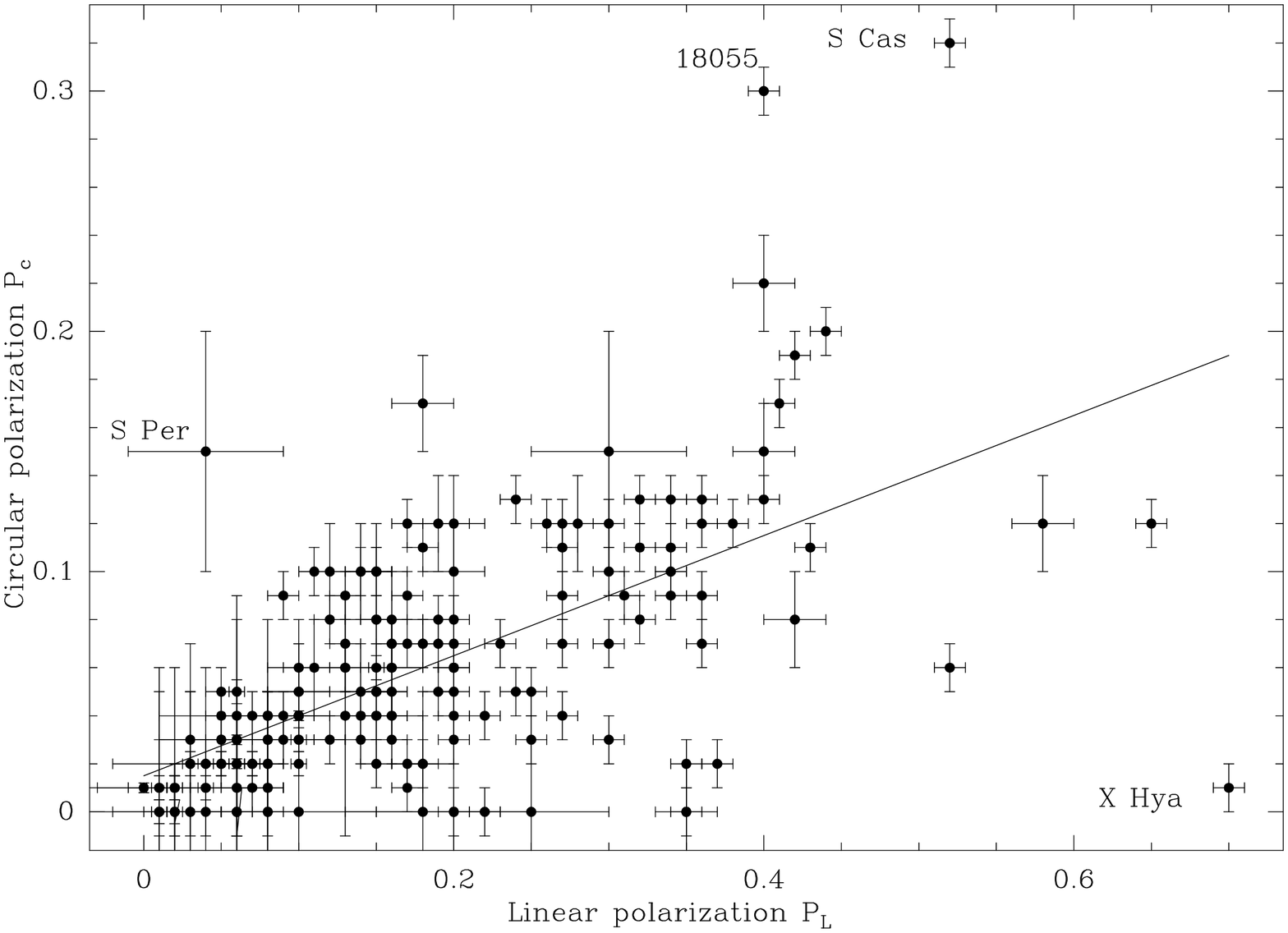}
\caption{Absolute fractional polarization $p_c$ versus $p_L$ for the whole sample and all maser components (wings + center). A regression fit is shown: $p_C \simeq 0.015 + 0.25 p_L$}
\label{ }
\end{center}
\end{figure}

We may look for any correlation between $p_C$ and $p_L$.  As shown in Figs. 2 and 3, it first appears that in all cases $p_L$ is larger than $p_C$ as is predicted by all models. More precisely, $p_C$ is noticeable if and only if $p_L$ is high. If we plot the values of  $p_C$ and $p_L$ derived for all maser components in this work (Fig. 4) and make a regression fit to our data we obtain $p_C \simeq 0.015 + 0.25 ~p_L$. The circular polarization level tends to vary approximately linearly with $p_L$ in agreement with the fact that no $p_C$ is detected towards sources with a marginal $p_L$ detection. Note that in Fig. 4, four objects, S Per, IRAS18055-1433, S Cas and $\chi$ Hya, do not follow the same general trend observed for the rest of the sample. (The case of the Supergiant S Per, however, is an exception as it exhibits substantial $p_C$ while $p_L<p_C$.)  This observation may in fact favour Watson's model. It must be stressed again that the beam-averaged polarization that we measure makes any conclusion uncertain. In fact, due to this averaging, we should observe no correlation at all, even if such one would exist !
  
 As mentioned earlier (see Section 3.2) we observe relatively high circular polarization rates in several stars (${<p_C>}_{Mira} \simeq 9$\%, ${<p_C>}_{SR} \simeq 5$\%). These values are larger than those predicted from Watson's model (e.g. Nedoluha \& Watson 1994). We also have not been able to find any correlation of $p_C$ with total intensity. Finally, although we adopt the Zeeman case to derive the magnetic field strength (see Section 4.5), we cannot conclude firmly from present observations which maser theory prevails for SiO emission. 

\subsection{Magnetic field in AGB stars}

A better knowledge of the stellar magnetic field strength is crucial to understand the last stages in the life of an Asymptotic Giant Branch (hereafter AGB) star. These stages are characterized by a high mass loss process driven by the radiation pressure; they are also influenced by the magnetic field (Palen \& Fix 2000, Blackman, Frank \& Welch 2001, and references therein). A strong magnetic field may rule the mass loss geometry; in particular, it could be the cause of a higher or lower mass-loss rate in the equatorial plane (Soker 2002), and thus determine the global shaping of these objects. But, as the direct dynamical effect of the magnetic activity is much lower than that of the wind (although in local spots the magnetic field can be dominant), the role of the magnetic field might be indirect. Moreover, the observed high mass loss rates are hardly explained by a single process and need a combination of several factors such as rotation, the presence of a companion (binary stars, with our without common envelope, exhibiting mass transfer or tidal effects are common; Soker 1997) and a magnetic field. During its quick transition to the PN stage, the AGB star will completely change its geometry: the quasi-spherical object becomes axisymmetrical, point-like symmetrical or even shows higher order symmetries (Johnson \& Jones 1991, Sahai \& Trauger 1998, Balick \& Frank 2002). The classical  or generalized {\em Interacting Stellar Winds} (hereafter ISW or GISW) models (Kwok 2000, Soker \& Livio 1989, Morris 1987) try to explain this shaping, but have serious difficulties in producing complicated structures with peculiar jets or ansae (e.g., CRL 2688, Delamarter 2000) and do not fully address the origin of the wind. Furthermore, recent X-ray studies with the Chandra satellite do not completely agree with GISW predictions for temperatures (Guerrero \etal 2001). 

Some recent studies tend to demonstrate the importance of the magnetic field in evolved objects. Bujarrabal \etal (2001) show that for 80\% of the PPNe in their sample the fast molecular flows have too high momenta to be powered by radiation pressure only (1000 times larger in some cases) what may be explained by magnetic field. Moreover, X-ray emission found in evolved stars (e.g. H\"unch, Schmitt \& Schr\"oder 1998) may indicate the presence of a hot corona that possibly results from magnetic activity. Very recently, magnetic field was discovered for the first time in central stars of PN (Jordan \etal 2005) and estimated to kiloGauss, much stronger than what we find here from our SiO data in QX Pup.

New models involving the magnetic field have been developed trying to explain the morphology changes of an object during its transition from the AGB stage to the PN stage; B plays the role of a catalyst and of a collimating agent. The most simple models are based on a moderately weak magnetic field alone ($B \simeq 1$ Gauss at the stellar surface, a few $10^{13}$ cm, i.e. at a radiis of a few AU, Soker 1998). The influence of B is stressed by the work of Smith \etal (2001) and Greaves (2002) in VY CMa. But the role of B can only be decisive when its energy density is greater than the radiative pressure, i.e. when B is greater than around 10 G close to the stellar surface in the SiO region (see Soker \& Zaobi, 2002). Arguing that such a strong field may be very unusual, Balick \& Frank (2002) explain that B alone cannot produce the observed structures, and a combination of several factors has thus to be considered (rotation, magnetic field and presence of a companion).

Soker \& Harpaz (1992) first proposed a model with a weak magnetic field ($\leq 1$ G) and included a slow rotation together with the presence of a companion to transform the envelope (and lead for example to the peculiar geometry observed in NGC 6826 or NGC 6543). Even if the star were not binary, the influence of B  is probably important locally (Palen \& Fix 2000). A significant magnetic field can form cold spots on the star's surface and a slow rotation of the star can then increase the field strength to build up a dipolar magnetic field varying as $1/r^{3}$ (Matt \etal 2000); such a field is  stronger at the equator and may thus lead to an axisymmetrical mass loss. 

The main argument against the dominant influence of the magnetic field on the shaping of the circumstellar envelope is that a strong field seems to be necessary to dominate the dynamics of the gas. However, several authors (Pascoli 1985, 1992, 1997, Chevalier \& Luo 1994, Garc\'{\i}a-Segura 1997, Gurzadyan 1997, Delamarter 2000) have demonstrated the strong influence of a reasonable toro\"{\i}dal magnetic field embedded in the normal radiation-driven stellar wind ({\em Magnetic Wind Bubble} theory, hereafter {\em MWB}). This field has a strength between a few Gauss and a few 10 Gauss at a few stellar radii (the SiO region is believed to be at $\sim 10^{14}$ cm or $\sim 7-10$ AU), varies as $1/r^2$, then as $1/r$ at larger radii; therefore $B\sim1$ mG at $10^{16,17}$ cm or  700-7000 AU. These results are confirmed by the simulations of Garc\'{\i}a-Segura, Lopez \& Franco (2001) for the PN He 2-90. Even if the origin of the wind is not explained by these models, it seems clear that a magnetic field is essential to generate fast collimated outflows (Kastner \etal 2003). 

There are many models of magnetic jet production and collimation and some, or all of them, are applicable to various star geometries. One most interesting study was performed by Blackman, Frank \& Welch (2001) in which the magnetic field emerges from the AGB stellar core and the resulting 1G field helps to collimate the radiation-driven wind or a stronger, more anisotropic, magnetically driven wind.

\subsection{Magnetic field in our sample}

The exact interpretation, in terms of magnetic field, of our observations depends on the adopted specific SiO maser model (see Sections 4.2 and 4.3). From the current knowledge of the strength of the magnetic field in the OH and \water layers we expect  $B_{//}$ of a few Gauss at least in the SiO maser region (see also Kemball \& Diamond 1997), i.e. at 5-10 AU from the central object. This tends to invalidate Watson's model, and furthermore tends to agree with a field varying in $r^{-1,2}$ as predicted by Elitzur's model. However, Vlemmings \etal (2005) measured the circular polarization of the \water maser emission in a few evolved stars with the VLBA observations and showed that the magnetic field is either a solar-type field (with a $r^{-2}$ field strength dependence) or a dipole magnetic field (with a $r^{-3}$ dependence) in their sample. 

In the following, we decide to use Elitzur's theory (Zeeman case) to infer magnetic field strength from the circular polarization levels. From equation (12), we thus calculate the mean value of the magnetic field $B_{//}$ for each star and give results in Table 3.

For our sample B$_{//}$ is between 0 and 20 Gauss, with a mean value of 3.5 G. This value combined with the strength of the field in more outer layers of the envelope (OH and \water masers) agrees with a B field variation law in $1/r$, closer to Elitzur's model. As explained in the Introduction and in Section 4.4, B alone can be the main agent to shape the circumstellar envelope if its value is larger than around 10 Gauss. This means that only S Cas, RU Aur, IRAS 18055-1433 and IRAS 18158-1527 may have a {\em magnetic field ruled geometry} ($B> 10$ G in these objects). The rest of our sample shows that B is sufficiently strong to be dominant at this stage of the AGB star evolution, but it should be associated with rotation and the presence of a companion as suggested in models mentioned in Section 4.4. 
Despite our B measurements are beam averaged, they suggest in many cases that they are not too much (not orders of magnitude) below the critical value; local B values may exceed in many cases the critical value, and therefore participate in the shaping of the AGB
envelopes. Our estimated values of B are consistent with the {\em MWB} theory (toro\"{\i}dal magnetic field) or the model of Blackman, Frank \& Welch (2001). 

\begin{table} [h]
 \caption{ \label{table} Average magnetic field strengths derived from $p_C$.}
 {\begin{tabular}{lc} \hline
 {\bf Source} & $B_{//}$ (G) \\ \hline
 IRAS18055-1433 & 4.6-13.9 \\
 IRAS18158-1527 & 3.7-20.0 \\
 IRAS18204-1344 & 0-3.2 \\
 W And & 0.4-5.9 \\
 AU Aur & 0.9-4.2 \\
 NV Aur & 1.9-4.6 \\
 R Aur & 0-6.0 \\
 RU Aur & 0-10.2 \\
 TX Cam & 0.4-2.8 \\
 V Cam & 0-1.9 \\
 R Cnc & 0.4-5.6 \\
 W Cnc & 1.4-6.0 \\
 VY CMa & 0-1.9 \\
 S CMi & 0-3.2 \\
R Cas &  0.9-2.8 \\
S Cas & 0-14.9 \\
T Cas & 0.9-5.1 \\
T Cep & 1.4-1.9 \\
R Com & 0.9-3.2 \\
S Crb & 0-3.2 \\
R Crt & 0-3.7 \\
$\chi$ Cyg & 0-8.8 \\
UX Cyg & 0.4-1.9 \\
R Hya & 0.9-4.6 \\
W Hya & 0.9-2.8 \\
X Hya & 0.4-1.4 \\
R Leo & 4.2-4.6 \\
W Leo & 0-3.7 \\
R LMi & 0-5.6 \\
T Lep & 1.4-5.1 \\
RS Lib & 0.4-4.6 \\
Ap Lyn & 0.9-5.6 \\
U Lyn & 0-6.0 \\
GX Mon & 0.4-5.6 \\
SY Mon & 2.3-5.6 \\
V Mon & 2.3 \\
U Ori & 0.9-6.0 \\
RR Per & 1.4-5.1 \\
S Per & 0-7.0 \\
QX Pup & 0-7.9 \\
Z Pup & 3.2 \\ 
VX Sgr & 0-1.4 \\
AH Sco & 0-2.3 \\
RR Sco & 0-4.6 \\
R Ser & 3.2 \\
S Ser & 0-1.9 \\
WX Ser & 1.9-5.6 \\
IK Tau & 1.9-6.0 \\
R Tau & 2.8-9.3 \\
RX Tau & 0.4-2.8 \\
R Tri & 2.3 \\
R UMi & 0 \\
S UMi & 0-0.9 \\
R Vir & 4.6-7.0 \\
RS Vir & 0-5.1 \\
RT Vir & 0-5.6 \\
S Vir & 1.4-2.3 \\ \hline 
  \end{tabular}}
\end{table}

From Elitzur (1996) and our measurements ${<p_C>}_{Mira} \simeq 9$\%, ${<p_C>}_{SR} \simeq 5$\%, we can estimate (see Eq. (9) in Sect. 4.2) that $<x_B>$ is around 0.017 and $9.4$ $10^{-3}$ respectively for Mira-type objects and semi-regular variables. Moreover, according to Elitzur (1996, see Fig.2), there is no stationary physical solutions for propagation at ${\sin}^2 \theta < \frac {1}{3}$ (i.e. at ${\sin}^2 \theta < \frac {1}{3}$ the radiation is not polarized). As $x_B$ is small ($<0.02$), from Fig.2 of Elitzur (1996) we can estimate the volume of phase space in which propagation of linear polarization in a maser is possible or not ($\theta > 35.3^{\circ}$ or $\theta < 35.3^{\circ}$); we then calculate that the probability for a random magnetic axis to be aligned with a given direction (our line of sight) is better than $35.3^{\circ}$. Our present estimate is that around 18 \%. Therefore, Elitzur's model predicts that 18.4 \% of the SiO 86 GHz masers should not be linearly polarized, because such polarized masers cannot propagate if the magnetic field, although weak, is closer than $35.3^{\circ}$ to the line of sight (propagation direction).  Hence non polarized maser emissions do not imply no or weak magnetic field. In our sample, roughly 13 \% of the SiO maser components have no detectable or very weak ($<3\%$) polarization.

We looked without success for a possible correlation between the polarization rates and physical parameters such as the known envelope asymmetry, the presence of SiO maser high velocity linewings (see Herpin \etal 1998), or the mass loss rate. If the magnetic field plays an important role in the shaping of the object, one may expect to find a relationship between the strength of $B_{//}$ (thus $p_C$) and the geometry of the object. Unfortunately, no trend is clearly found in our data. Nevertheless it is known that radiative pressure is driving the wind in AGB objects and it is thus not surprising that we find no correlation between the B strength and a known asymmetry in our sample. Of course, our stellar sample would require new observations with sufficient spatial resolution (VLBI) to confirm the present results; the same type of study should also be conducted toward several Proto-PN and PN objects. 
 
\section{Conclusion}

We have made a study of the SiO maser polarization in a representative sample of evolved stars, simultaneously measuring, for the first time, the 4 Stokes parameters. From our measurements we derive the circular and linear polarization levels and shows that, due to the beam averaging of our polarization measurements, we cannot firmly discriminate between the two dominant theories of SiO maser emission. In particular, VLBI observations of our source sample are absolutely necessary to distinguish between Zeeman or non-Zeeman theories. Nevertheless, the magnetic field strength was derived assuming Elitzur's model. $B_{\\}$ varies between 0 and 20 Gauss, with a mean value of 3.5 G. As a consequence, we suggest that the magnetic field plays a significant role in the evolution of these objects. Within the frame of the Zeeman theory the magnetic field could shape or even collimate the gas layers surrounding the AGB objects. Emission from Mira-type objects clearly tends to have a higher linear  ( ${<p_L>}_{Mira} \simeq 30$\%, ${<p_L>}_{SR} \simeq 11$\%) and circular polarization  (${<p_C>}_{Mira} \simeq 9$\%, ${<p_C>}_{SR} \simeq 5$\%). Basically, if there is a real correlation between $p_C$ and the strength of the magnetic field, this trend may indicate that the magnetic field may be  stronger in Mira objects than in Semi-Regular variables (at least in the inner layers of the circumstellar envelope).

To better understand the mechanisms at work with the magnetic field, complementary studies have to be conducted and in particular the presence of a companion has to be investigated in a large sample of objects. Of course VLBI maps of the magnetic field in these stars are essential.  
Another important objective is to investigate the evolution of the magnetic field and its influence during the transition from the AGB star phase to the PN stage.

\acknowledgements{The authors are grateful to M. Elitzur for reading and commenting on this paper. We also thank W.D. Watson for his useful comments and suggestions. The authors are indebted to the staff of the IRAM 30m telescope who most efficiently helped during the observations and to R. Mauersberger who closely followed part of these observations. Finally, we also thank the referee for several useful comments.}

\Online
  \addtocounter{figure}{-3}

\begin{figure*} [ht] 
  \begin{center} 
     \epsfxsize=13cm 
     \epsfbox{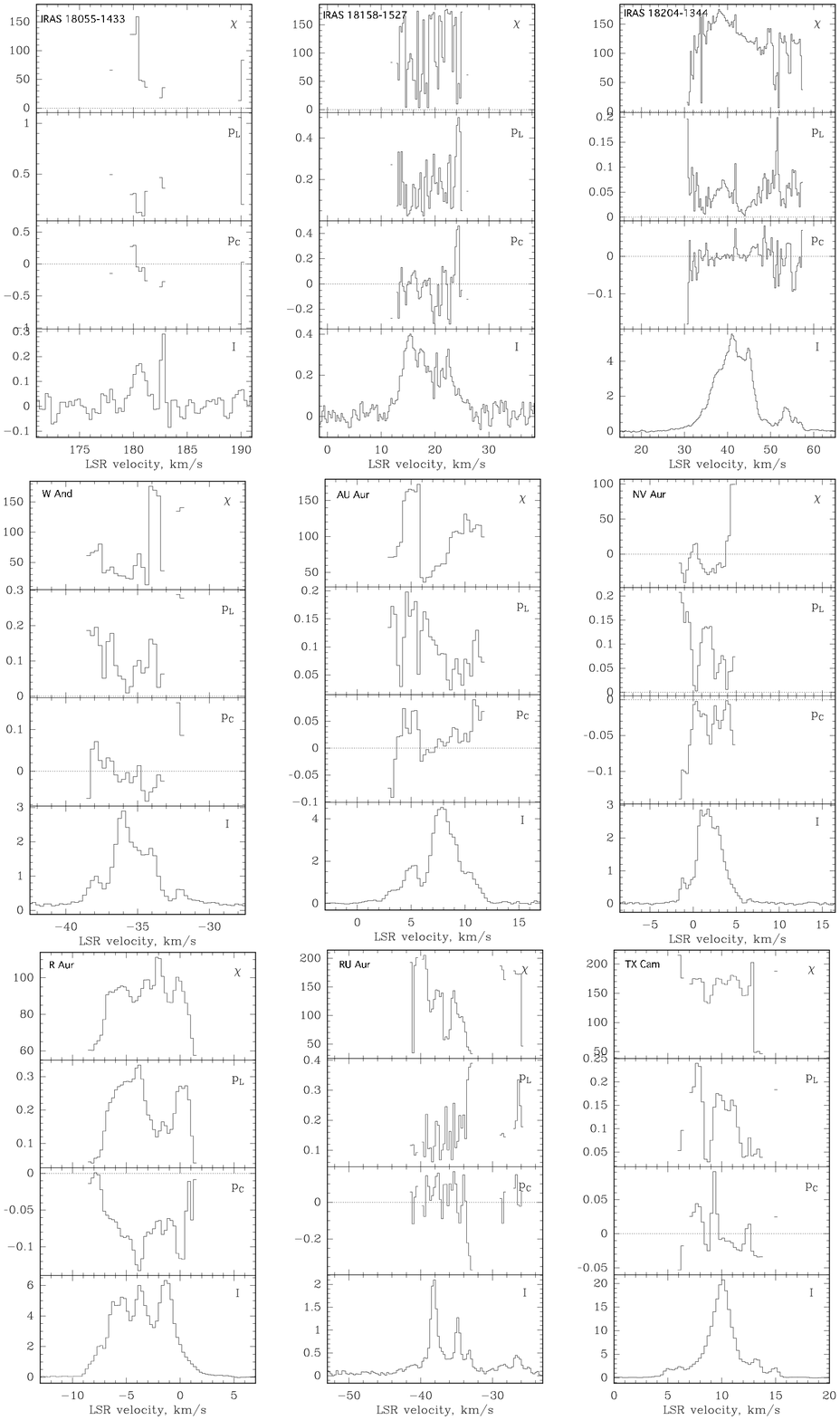} 
  \end{center} 
  \caption []{Position angle of polarization ($\chi$) in degrees, linear ($p_L$) and circular ($p_C$) polarization levels and intensity ($I= Ta^{\star}$ in Kelvin; the conversion factor is 6 Jy$/$K) for the SiO emission observed in several stars. } 
  \label{} 
\end{figure*} 
  \addtocounter{figure}{-1}

\begin{figure*} [ht] 
  \begin{center} 
     \epsfxsize=13cm 
     \epsfbox{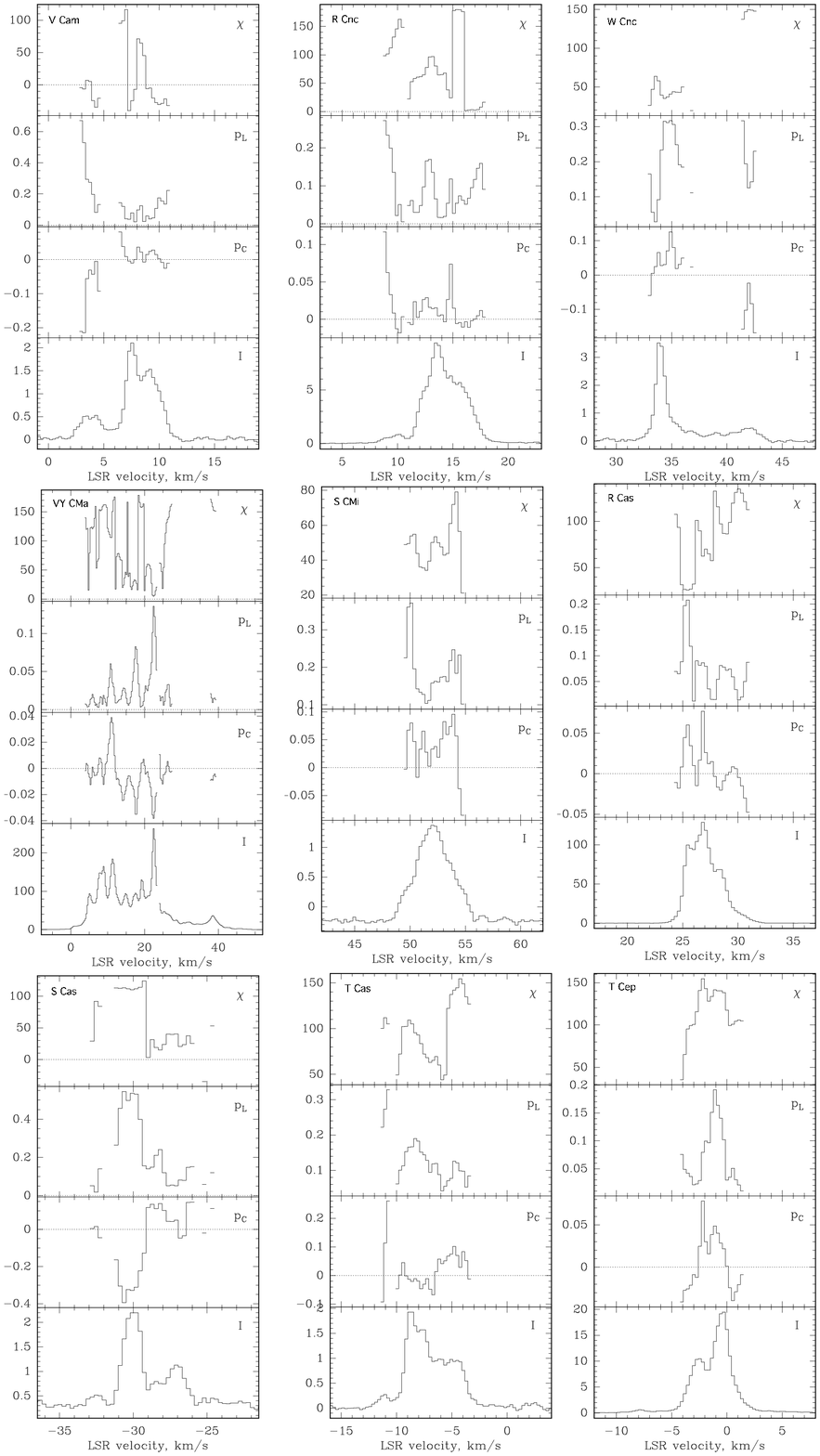} 
  \end{center} 
  \caption []{(-continued). Position angle of polarization ($\chi$) in degrees, linear ($p_L$) and circular ($p_C$) polarization levels and intensity ($I= Ta^{\star}$ in Kelvin; the conversion factor is 6 Jy$/$K) for the SiO emission observed in several stars. } 
  \label{} 
\end{figure*} 
  \addtocounter{figure}{-1}

\begin{figure*} [ht] 
  \begin{center} 
     \epsfxsize=13cm 
     \epsfbox{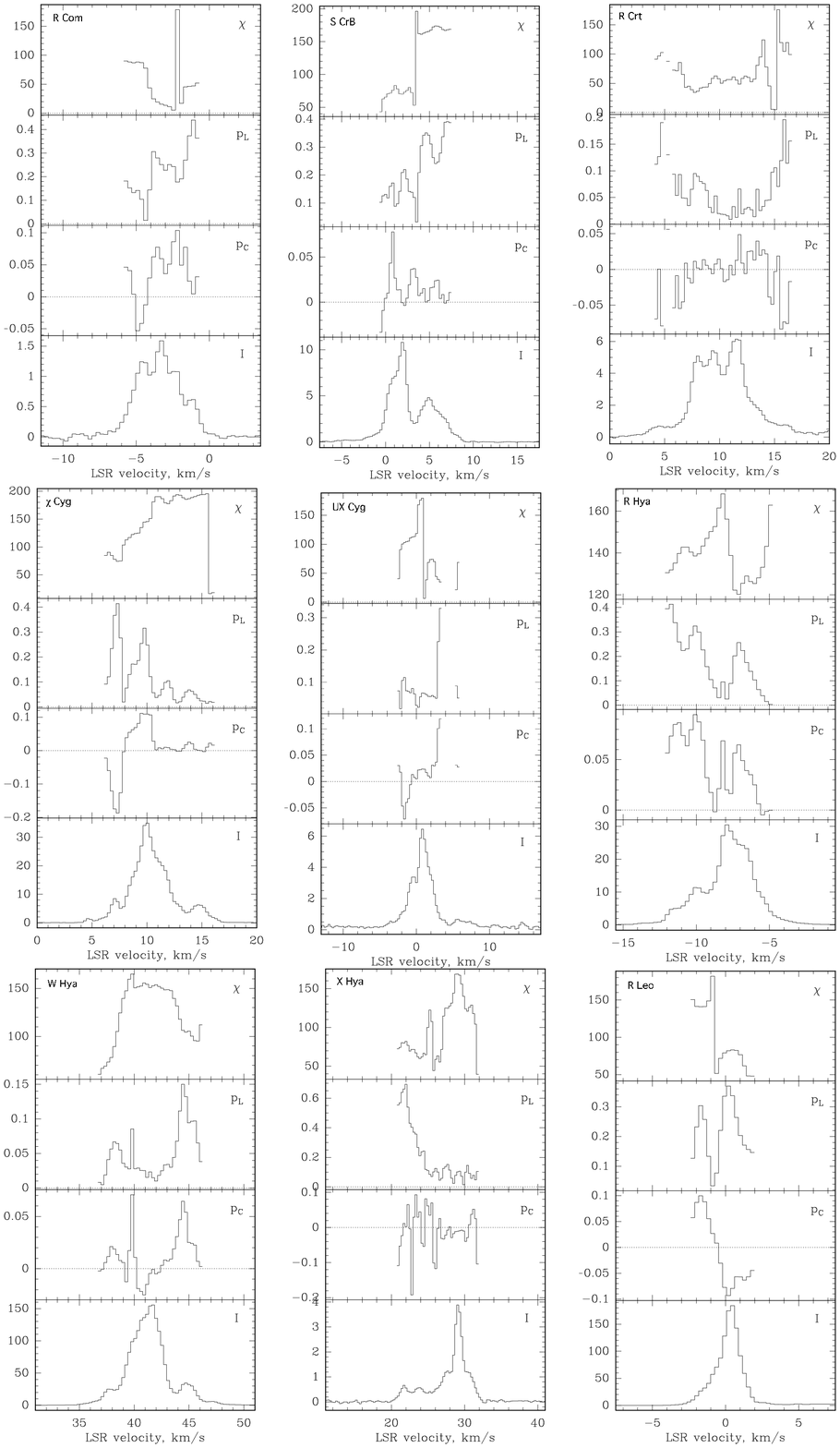} 
  \end{center} 
  \caption []{(-continued). Position angle of polarization ($\chi$) in degrees, linear ($p_L$) and circular ($p_C$) polarization levels and intensity ($I= Ta^{\star}$ in Kelvin; the conversion factor is 6 Jy$/$K) for the SiO emission observed in several stars. } 
  \label{} 
\end{figure*} 
  \addtocounter{figure}{-1}

\begin{figure*} [ht] 
  \begin{center} 
     \epsfxsize=13cm 
     \epsfbox{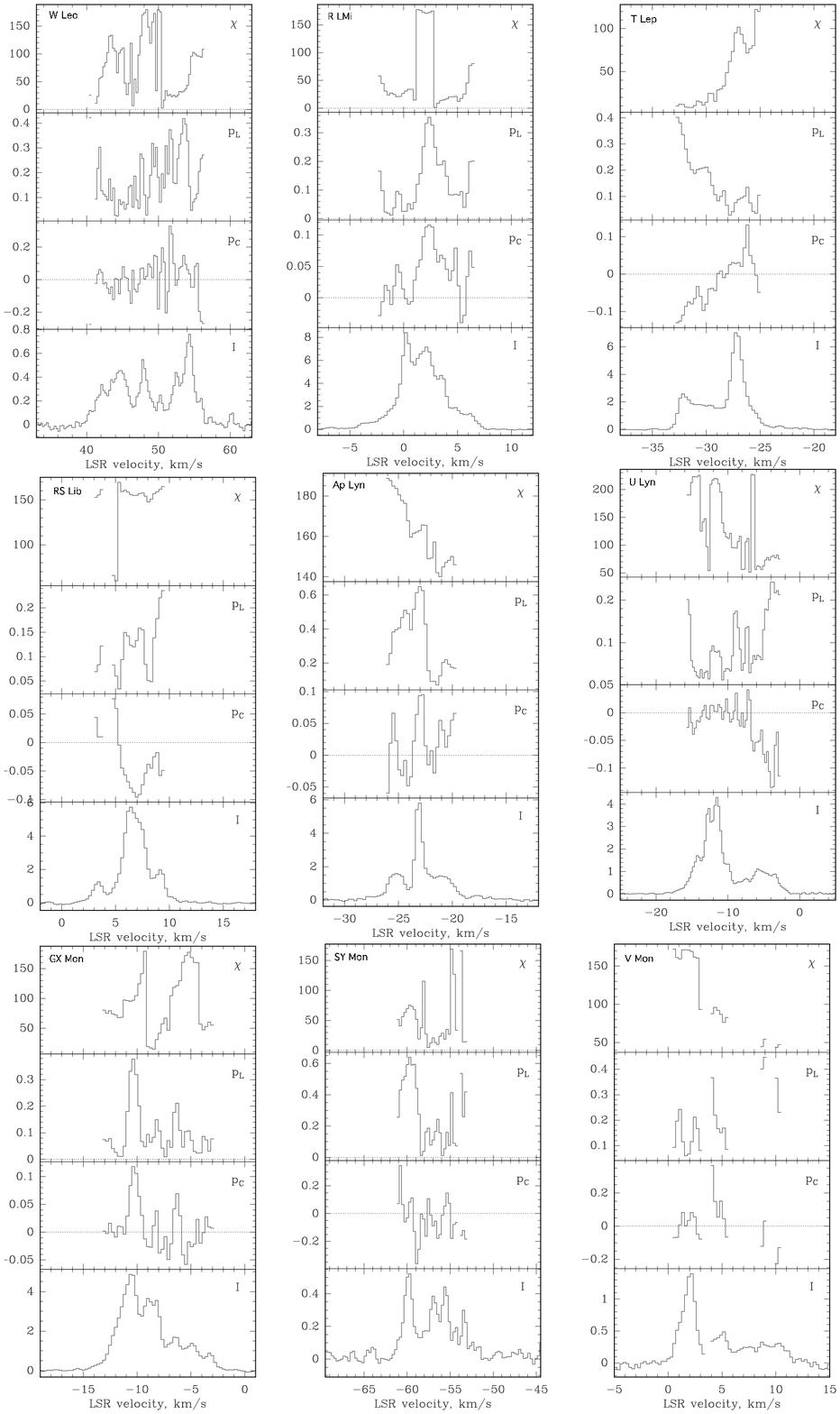} 
  \end{center} 
  \caption []{(-continued). Position angle of polarization ($\chi$) in degrees, linear ($p_L$) and circular ($p_C$) polarization levels and intensity ($I= Ta^{\star}$ in Kelvin; the conversion factor is 6 Jy$/$K) for the SiO emission observed in several stars. } 
  \label{} 
\end{figure*} 
  \addtocounter{figure}{-1}

\begin{figure*} [ht] 
  \begin{center} 
     \epsfxsize=13cm 
     \epsfbox{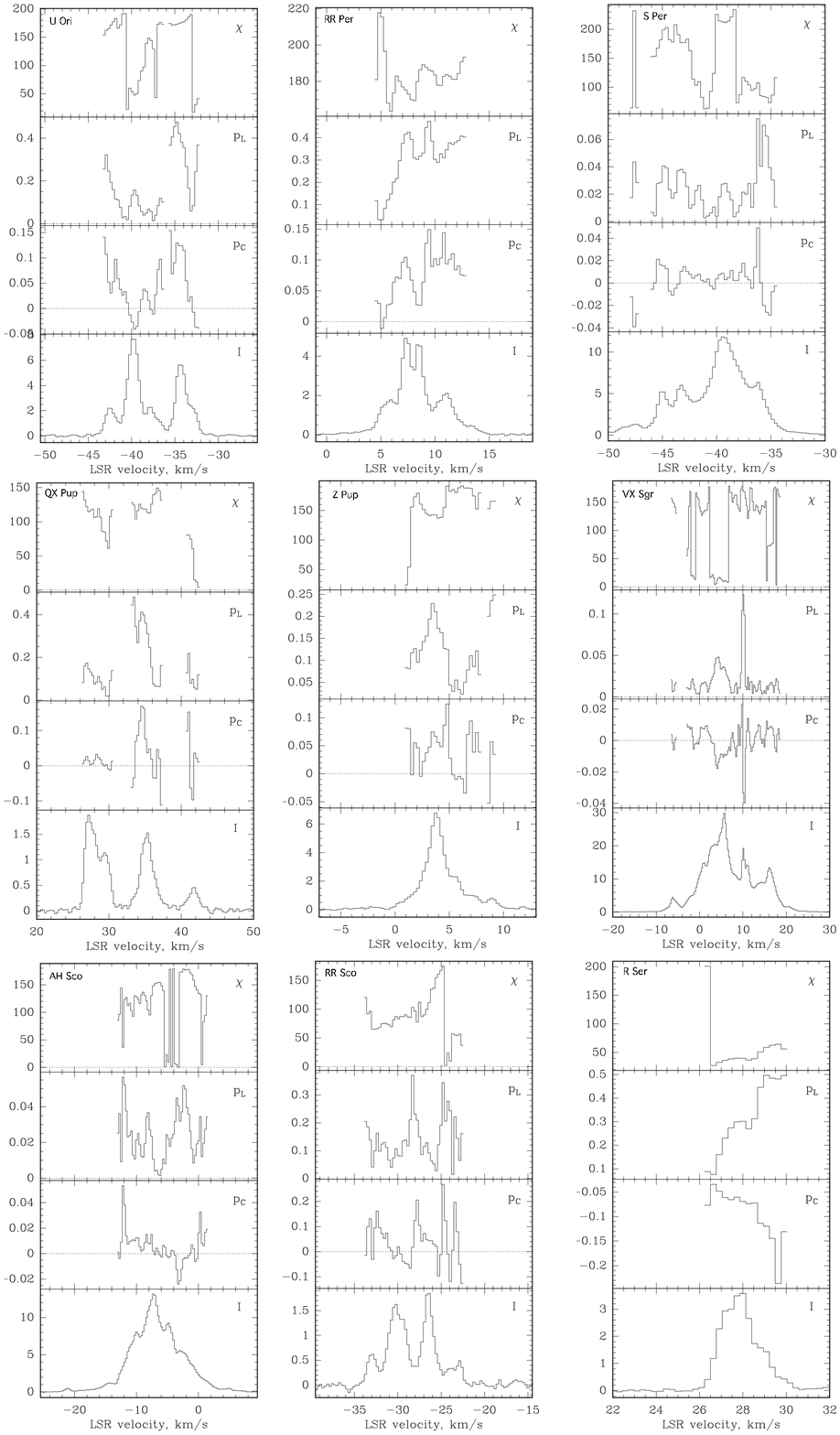} 
  \end{center} 
  \caption []{(-continued). Position angle of polarization ($\chi$) in degrees, linear ($p_L$) and circular ($p_C$) polarization levels and intensity ($I= Ta^{\star}$ in Kelvin; the conversion factor is 6 Jy$/$K) for the SiO emission observed in several stars. } 
  \label{} 
\end{figure*} 
  \addtocounter{figure}{-1}

\begin{figure*} [ht] 
  \begin{center} 
     \epsfxsize=13cm 
     \epsfbox{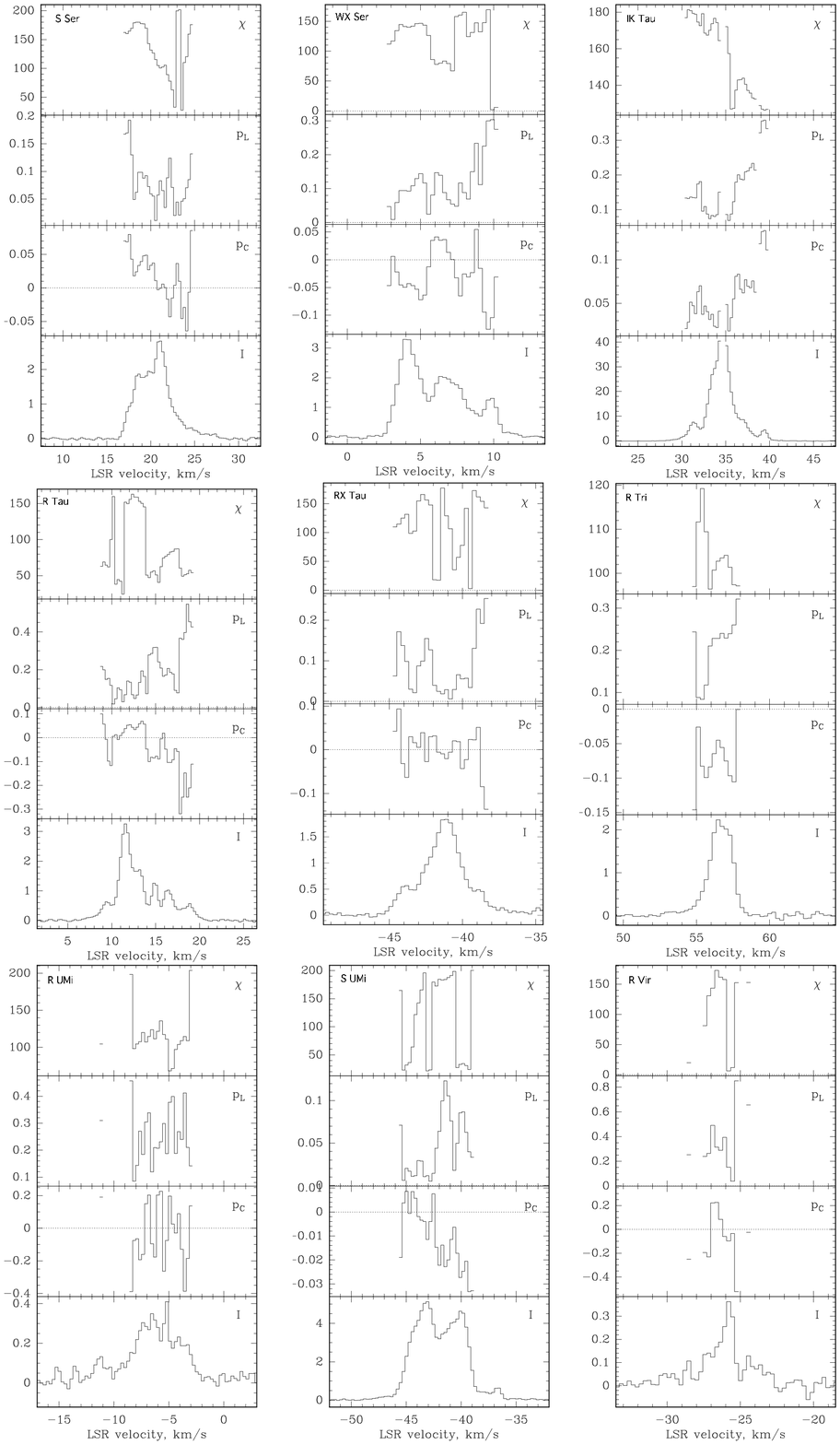} 
  \end{center} 
  \caption []{(-continued). Position angle of polarization ($\chi$) in degrees, linear ($p_L$) and circular ($p_C$) polarization levels and intensity ($I= Ta^{\star}$ in Kelvin; the conversion factor is 6 Jy$/$K) for the SiO emission observed in several stars. } 
  \label{} 
\end{figure*} 
  \addtocounter{figure}{-1}

\begin{figure*} [ht] 
  \begin{center} 
     \epsfxsize=13cm 
     \epsfbox{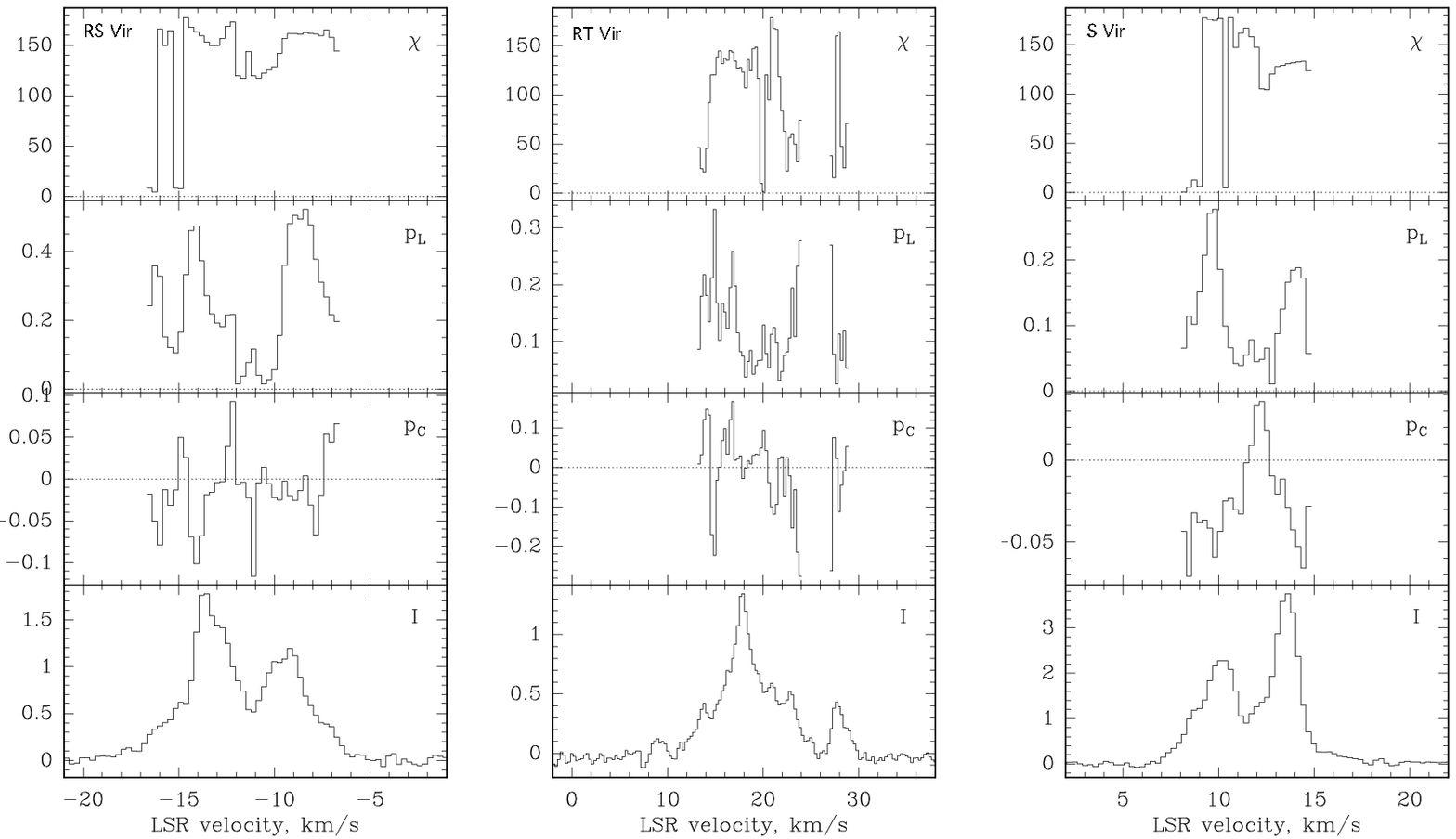} 
  \end{center} 
  \caption []{(-continued). Position angle of polarization ($\chi$) in degrees, linear ($p_L$) and circular ($p_C$) polarization levels and intensity ($I= Ta^{\star}$ in Kelvin; the conversion factor is 6 Jy$/$K) for the SiO emission observed in several stars. } 
  \label{} 
\end{figure*} 


\begin{thebibliography}{}
\bibitem[]{}
Balick B., Frank A. 2002, ARAA 40, 439
\bibitem[]{}
Barvainis R., McIntosh G., Read Predmore C. 1987, \textit{Nature} 329, 613
\bibitem[]{}
Blackman E.G., Frank A., Welch C. 2001, ApJ 546, 288
\bibitem[]{}
Bujarrabal V., Castro-Carrizo A., Alcolea J., S\'anchez Contreras C. 2001, A\&A 377, 868
\bibitem[]{}
Bujarrabal V. 2003, Mass-losing pulsating stars and their circumstellar matter, edited by Y. Nakada, M. Honma and M. Seki. Astrophysics and Space Science Library, Vol. 283, Dordrecht: Kluwer Academic Publishers, p. 275 - 282
\bibitem[]{}
Chevalier R.A., Luo D. 1994, ApJ 435, 815
\bibitem[]{}
Cotton W.D., Mennesson B., Diamond P.J. \etal 2004, A\&A 414, 275
\bibitem[]{}
Delamarter G.R. 2000, Ph-D thesis, University of Rochester
\bibitem[]{}
Elitzur M. 1994, ApJ 422, 751
\bibitem[]{}
Elitzur M. 1996, ApJ 457, 415
\bibitem[]{}
Elitzur M. 1998, ApJ 504, 390
\bibitem[]{}
Elitzur M. 2002, Astrophysical spectropolarimetry, Proceedings of the XII Canary Islands Winter School of Astrophysics, edited by J. Trujillo-Bueno, F. Moreno-Insertis, and F. S‡nchez. Cambridge, UK: Cambridge University Press,  p. 225 - 264
\bibitem[]{}
Fiebig D., G\"usten R. 1989, A\&A 214, 333
\bibitem[]{}
Garc\'{\i}a-Segura G. 1997, ApJ 489, L189
\bibitem[]{}
Garc\'{\i}a-Segura G, Lopez J.A. \& Franco J. 2001, ApJ 560, 928
\bibitem[]{}
Glenn J., Jewell P.R., Fourre R., Miaja L. 2003, ApJ 588, 478
\bibitem[]{}
Goldreich P., Keeley D.A., Kwan J.Y. 1973, ApJ 179, 111
\bibitem[]{}
Greaves J.S. 2002, A\&A 392, L1
\bibitem[]{}
Guerrero M.A., Chu Y., Gruendl R.A. \etal 2001, ApJ 553, L55
\bibitem[]{}
Gurzadyan G.A. 1997, in {\em The Physics and Dynamics of Planetary Nebulae}, Berlin/Heidelberg/New York: Springe-Verlag
\bibitem[]{}
Herpin F., Baudry A., Alcolea J., Cernicharo J. 1998, A\&A 334, 1037
\bibitem[]{}
Herwig F. 2003, Planetary Nebulae: Their Evolution and Role in the Universe, Proceedings of the 209th Symposium of the International Astronomical Union, edited by Sun Kwok, Michael Dopita, and Ralph Sutherland,  p.61
\bibitem[]{}
H\"unch M., Schmitt J.H.M.M., Scrh\"oder K.-P. 1998, A\&A 330, 225
\bibitem[]{}
Jordan S., Werner K., O'Toole S.J. 2005, A\&A 432, 273
\bibitem[]{}
Johnson D.R., Clark F.O. 1975, ApJ 197, L69
\bibitem[]{}
Johnson J.J., Jones T.J. 1991, AJ 101, 1735
\bibitem[]{}
Kastner J.H., Balick B., Blackman E.G. \etal 2003, ApJ 591, L37
\bibitem[]{}
Kemball A.J., Diamond P.J. 1997 ApJ 481, L111
\bibitem[]{}
Kwok S. 2000, in {\em The origin and Evolution of Planetary Nebulae}, Cambridge Astrophysics Series 31, Cambridge University Press edition
\bibitem[]{}
Loup C., Forveille T., Omont A., PAul J.-F. 1993, A\&ASS 99, 291
\bibitem[]{}
McIntosh G.C., Predmore C.R., Moran J.M. \etal 1989, ApJ 337, 934
\bibitem[]{}
McIntosh G.C., Predmore C.R., Patel N.A. 1994, ApJ 428, L29
\bibitem[]{}
Matt S., Balick B., Winglee R., Goodson A. 2000, ApJ 545, 965
\bibitem[]{}
Monnier J.D., Millan-Gabet R., Tuthill P.G. \etal 2004, ApJ 605, 436 
\bibitem[]{}
Morris M. 1987, PASP 99, 1115
\bibitem[]{}
Nedoluha G.E., Watson W.D. 1994, ApJ 423, 394
\bibitem[]{}
Palen S., Fix J.D. 2000, ApJ 531, 391
\bibitem[]{}
Pascoli G. 1985, A\&A 147, 257
\bibitem[]{}
Pascoli G. 1992, PASP 104, 350
\bibitem[]{}
Pascoli G. 1997, ApJ 489, 946
\bibitem[]{}
Plambeck R.L., Wright M.C.H., Rao R. 2003, ApJ 594, 911
\bibitem[]{}
Rowan-Robinson M., Harris S. 1983, MNRAS 202, 767
\bibitem[]{}
Sahai R., Trauger J.T. 1998, AJ 116, 1357
\bibitem[]{}
Smith N., Humphreys R.M., Davidson K. \etal 2001, ApJ 121, 1111
\bibitem[]{}
Soker N., Livio M. 1989, ApJ 339, 268
\bibitem[]{}
Soker N., Harpaz A. 1992, PASP 104, 923
\bibitem[]{}
Soker N. 1997, ApJS 112, 487
\bibitem[]{}
Soker N. 1998, MNRAS 299, 1242
\bibitem[]{}
Soker N., Zoabi E. 2002, MNRAS 329, 204
\bibitem[]{}
Soker N. 2002, MNRAS 337, 1038
\bibitem[]{}
Szymczak M., Cohen R.J. 1997, MNRAS 288, 945
\bibitem[]{} 
Thum C., Wiesemeyer H., Morris D., Navarro S., Torres M. 2003 {\em Proc. S.P.I.E.}, 4843, 272
\bibitem[]{}
Troland T.H., Heiles C., Johnson D.R., Clark F.O. 1979, ApJ 232, 143
\bibitem[]{}
van der Veen W.E.C.J., Habing H.J. 1988, A\&A 194, 125
\bibitem[]{}
Vlemmings W., Diamond P.J., van Langevelde H.J. 2001, A\&A 375, L1
\bibitem[]{}
Vlemmings W., van Langevelde H.J., Diamond P.J.  2005, A\&A 434, 1029
\bibitem[]{}
Wallin B.K., Watson W.D. 1997, ApJ 481, 832
\bibitem[]{}
Watson W.D., Wyld H.W. 2001, ApJL 558, 55
\bibitem[]{}
Wiebe D.S., Watson W.D. 1998, ApJL 503, 71
\bibitem[]{}
Wiesemeyer H., Thum C., Walmsley C.M. 2004, A\&A 428, 479
\end{thebibliography}
\end{document}